\documentstyle[prd,aps,eqsecnum,preprint,tighten,floats,psfig]{revtex}

\def\be{\begin{equation}}
\def\bea{\begin{eqnarray}}
\def\ee{\end{equation}}
\def\eea{\end{eqnarray}}
\def\d{\partial}

\begin{document}

\draft

\preprint{\vbox{\hfill UMN-D-01-7 \\
          \vbox{\hfill OHSTPY-HEP-T-01-016} \\
          \vbox{\hfill hep-th/0106193}
          \vbox{\vskip0.3in}
          }}

\title{Wave functions and properties of massive states
in three-dimensional supersymmetric Yang--Mills theory}

\author{John R. Hiller}
\address{Department of Physics, 
University of Minnesota Duluth, 
Duluth, MN 55812}

\author{Stephen Pinsky and Uwe Trittmann}
\address{Department of Physics, 
The Ohio State University, 
Columbus, OH 43210}

\date{\today}

\maketitle

\begin{abstract}
We apply supersymmetric discrete light-cone quantization (SDLCQ) 
to the study of supersymmetric Yang--Mills theory on
${\bf R} \times S^1 \times S^1$\@.  One of the compact
directions is chosen to be light-like and the other to be space-like.
Since the SDLCQ regularization explicitly preserves supersymmetry, 
this theory is totally finite, and thus we can solve for bound-state wave
functions and masses numerically without renormalizing.  
We present an overview of all the massive states of this theory, and 
we see that the spectrum divides into two distinct and
disjoint sectors. In one sector the SDLCQ approximation is only valid up to
intermediate coupling. There we find a well defined and 
well behaved set of states, and we present a detailed analysis of these 
states and their properties. In the other sector, which contains a completely
different set of states, we present a much more limited analysis for strong
coupling only.  We find that, while these state have a well defined 
spectrum, their masses grow with the transverse momentum cutoff.  
We present an overview of these states and their properties.
\end{abstract}

\narrowtext

\section{Introduction}

The properties of strongly coupled gauge theories
with a sufficient amount of supersymmetry can be analyzed in great 
detail~\cite{seibergwitten,seiberg,maldacena}. In particular, 
there are a number of supersymmetric gauge theories that are 
believed to be interconnected through a web of strong-weak 
coupling dualities. While these dualities provide a great 
deal of insight, there is still a need to study
the bound states of these theories directly 
and at any coupling.

It is well known that (1+1)-dimensional field theories {\em can} 
be solved from first principles via a straightforward application 
of discrete light-cone quantization (DLCQ)~\cite{pb85,bpp98}.
This includes a large class of supersymmetric gauge theories in two dimensions.
More recently a supersymmetric form of DLCQ (or `SDLCQ'),
which is known to preserve supersymmetry, has been 
developed~\cite{sakai95,klebhash,alp98a,alp98b,alpp98,alpp99,alp99}. 
We have recently shown that the SDLCQ algorithms can be extended to 
solve higher-dimensional theories~\cite{alp99b}. One important 
difference between two-dimensional and higher-dimensional theories 
is the phase diagram induced by variations in the gauge coupling. 
The spectrum of a (1+1)-dimensional gauge theory scales trivially 
with respect to the gauge coupling, while a theory in higher dimensions 
has the potential of exhibiting a complex phase structure,
which may include strong, intermediate, and weak coupling
regions~\cite{alp99b}. It is therefore interesting to study the 
properties of the bound states of gauge theories in $D \geq 3$ 
dimensions in all of these regions.

Towards this end, we apply SDLCQ to the study of 
three-dimensional SU($N_c$) ${\cal N}=1$ super-Yang--Mills theory 
compactified on the space-time ${\bf R} \times S^1 \times S^1$. 
This extends previous work~\cite{alp99b,hhlpt99} to better
numerical resolution and includes extraction of wave functions
for bound states at these higher resolutions.  We work in the
large-$N_c$ limit, with the light-cone coordinate $x^-$ and
transverse spatial coordinate $x_\perp$ compactified on the
circles $S^1$.  As is customary in DLCQ, we drop the longitudinal
zero modes ~\cite{sakai95,klebhash,alp98a,alp98b,alpp99,alp99b,%
ahlp99,hlpt99,dak93,kutasov93}. A review of dynamical and constrained 
zero modes can be found in~\cite{bpp98}. 

We are able to solve for bound states and their properties numerically by
diagonalizing the discretized light-cone supercharge. Since we do not break the
supersymmetry, the resulting spectrum is exactly supersymmetric. 
Study of the entire spectrum, which one obtains by a complete 
diagonalization of the entire Hamiltonian, shows that the spectrum 
breaks up into three distinct parts, which we will call the weak, 
intermediate, and strong-coupling regions. The weak-coupling region is
closely related to the dimensionally reduced theory, and we have discussed it
elsewhere~\cite{alp99b}. By the intermediate-coupling region, we 
refer to the low-mass portion of the spectrum. We will see that this region 
clearly separates from the high-mass region for couplings beyond 
weak coupling. We will also see that a standard DLCQ analysis of
these low-mass states is only possible at intermediate coupling. 
We will present a detailed analysis of these states and their 
properties in this region. We then look at the high-mass spectrum and 
show that it appears to behave as a strongly coupled spectrum.
Unfortunately, a detailed analysis of these states requires a 
complete diagonalization of the Hamiltonian and that is beyond 
the scope of this work.

The remainder of the paper is structured as follows.  In
Sec.~\ref{formulation}, we summarize the formulation of SU($N_c$)
${\cal N}=1$ super-Yang--Mills theory defined on the compactified
space-time ${\bf R} \times S^1 \times S^1$.  This includes explicit
expressions for the light-cone supercharges and their discretization
via SDLCQ. Also discussed are discrete symmetries of the theory that
are helpful in classifying the spectrum. In Sec.~\ref{numerical}, we
discuss the numerical  methods that we use and present an overview of
the full spectrum of the theory.  In Sec.~\ref{Icoupling}, we present a
detailed analysis of the states in the low-mass sector, where the SDLCQ
approximation is only valid up to intermediate coupling.  This
includes a study of the convergence of the states in both transverse
and longitudinal resolutions.  We present some of the
properties of these states and calculate transverse-momentum
distribution functions for some of these states.  In
Sec.~\ref{scoupling}, we consider the other distinct sector of
states.  We present the scaled mass spectrum as a function of the
inverse coupling and discuss the strong-coupling spectrum as a
function of the transverse momentum cutoff.  In Sec.~\ref{summary}, we
conclude by discussing the general properties of this theory and 
their implications. 
 
\section{Light-Cone Quantization and SDLCQ}
\label{formulation}

The action for three-dimensional ${\cal N}=1$ supersymmetric
Yang--Mills theory, in a space-time compactified on a scale
$L$ in the transverse direction $x_\perp=x^2$, is 
\be
S=\int d^2 x \int_0^L dx_\perp \mbox{tr}(-\frac{1}{4}F^{\mu\nu}F_{\mu\nu}+
{\rm i}{\bar\Psi}\gamma^\mu D_\mu\Psi)\,,
\ee
where $A_\mu$ and $\Psi$ are traceless $N_c\times N_c$ hermitian matrices
transforming in the adjoint representation of $SU(N_c)$, and we suppress
color indices.
We introduce light-cone coordinates $x^\pm=(x^0\pm x^1)/\sqrt{2}$
and choose to work in the light-cone gauge $A^+=0$.  With chiral projections
of the spinor $\Psi$ defined by
\be
\psi=\frac{1+\gamma^5}{2^{1/4}}\Psi\,,\qquad
\chi=\frac{1-\gamma^5}{2^{1/4}}\Psi\,,
\ee
and $\phi\equiv A^2$, the action becomes
\bea\label{action}
S&=&\int dx^+dx^- \int_0^L dx_\perp \mbox{tr}\left[\frac{1}{2}(\d_-A^-)^2+
(D_+\phi+\d_\perp A^-)\d_-\phi+ {\rm i}\psi D_+\psi+ \right.\nonumber \\
& &
\left. \hspace{15mm} +{\rm i}\chi\d_-\chi+\frac{{\rm i}}{\sqrt{2}}\psi
  D_\perp\chi+ \frac{{\rm i}}{\sqrt{2}}\chi D_\perp\psi \right]\,.
\eea
We choose the light-cone gauge because the non-dynamical fields $A^-$
and $\chi$ may be obtained explicitly from their equations of
motion, which are actually constraint equations in light-cone
coordinates.  These fields are then given in terms of the 
physical degrees of freedom $\phi$ and $\psi$ as 
\be
A^-=\frac{1}{\d_-^2}J=
\frac{1}{\d_-^2}\left(ig[\phi,\d_-\phi]+2g\psi\psi -\d_\perp \d\phi
\right)\,, \quad
\chi=-\frac{1}{\sqrt{2}\d_-}D_\perp\psi\,.\nonumber
\ee
The light-cone energy $P^-$ and momentum operators, $P^+$ and
$P^\perp$, become
\bea\label{moment}
P^+&=&\int dx^-\int_0^L dx_\perp\mbox{tr}\left[(\d_-\phi)^2+
{\rm i}\psi\d_-\psi\right],\\
P^-&=&\int dx^-\int_0^L dx_\perp\mbox{tr}
\left[-\frac{1}{2}J\frac{1}{\d_-^2}J-
            \frac{{\rm i}}{2}D_\perp\psi\frac{1}{\d_-}D_\perp\psi\right],\\
P^\perp &=&\int dx^-\int_0^L dx_\perp\mbox{tr}\left[\d_-\phi\d_\perp\phi+
            {\rm i}\psi\d_\perp\psi\right]\,.
\eea
The chiral projections of the light-cone supercharge, a
two-component Majorana spinor, are
\bea\label{sucharge}
Q^+&=&2^{1/4}\int dx^-\int_0^L dx_\perp\mbox{tr}\left[\phi\d_-\psi-\psi\d_-
                   \phi\right]\,,\\
\label{sucharge-}
Q^-&=&2^{3/4}\int dx^-\int_0^L dx_\perp\mbox{tr}\left[\d_\perp\phi\psi+
            g\left({\rm
i}[\phi,\d_-\phi]+2\psi\psi\right)\frac{1}{\d_-}\psi\right]\,.
\eea
At large $N_c$ the canonical \mbox{(anti-)}commutators for
the propagating fields are, at equal light-cone time $x^+$,
\begin{equation}
\left[\phi_{ij}(x^-,x_\perp),\d_-\phi_{kl}(y^-,y_\perp)\right]=
\left\{\psi_{ij}(x^-,x_\perp),\psi_{kl}(y^-,y_\perp)\right\}=
\frac{1}{2}\delta(x^- -y^-)\delta(x_\perp -y_\perp)\delta_{il}\delta_{jk}\,.
\label{comm}
\end{equation}
From these one can derive the supersymmetry algebra
\be
\{Q^+,Q^+\}=2\sqrt{2}P^+,\qquad \{Q^-,Q^-\}=2\sqrt{2}P^-,\qquad
\{Q^+,Q^-\}=-4P_\perp\,.
\label{superr}
\ee

We will only consider the sector where the total transverse
momentum is zero.  The other allowed sectors, given our compactification,
have transverse momentum $2\pi n/L$, with $n$ a nonzero integer.
In the sector with zero transverse momentum, $Q^+$ and $Q^-$
anticommute with each other, and the supersymmetry algebra
is equivalent to the ${\cal N}=(1,1)$ supersymmetry of
the two-dimensional theory obtained by dimensional 
reduction~\cite{sakai95}.  Also in this sector, the
mass squared operator is given by $2P^+P^-$.  The eigenvalue
problem for the bound states is then $2P^+P^-|M\rangle=M^2|M\rangle$,
with $|M\rangle$ expanded in a Fock basis diagonal in $P^+$
and $P^\perp$.

The expansions of the field operators in terms of creation and
annihilation operators for the Fock basis are
\bea
\lefteqn{
\phi_{ij}(0,x^-,x_\perp) =} & & \nonumber \\
& &
\qquad\frac{1}{\sqrt{2\pi L}}\sum_{n^{\perp} = -\infty}^{\infty}
\int_0^\infty
         \frac{dk^+}{\sqrt{2k^+}}\left[
         a_{ij}(k^+,n^{\perp})e^{-{\rm i}k^+x^- +{\rm i}
\frac{2 \pi n^{\perp}}{L} x_\perp}+
         a^\dagger_{ji}(k^+,n^{\perp})e^{{\rm i}k^+x^- -
{\rm i}\frac{2 \pi n^{\perp}}{L}  x_\perp}\right]\,,
\nonumber\\
\lefteqn{
\psi_{ij}(0,x^-,x_\perp) =} & & \nonumber \\
& & \qquad\frac{1}{2\sqrt{\pi L}}\sum_{n^{\perp}=-\infty}^{\infty}\int_0^\infty
         dk^+\left[b_{ij}(k^+,n^{\perp})e^{-{\rm i}k^+x^- +
{\rm i}\frac{2 \pi n^{\perp}}{L} x_\perp}+
         b^\dagger_{ji}(k^+,n^\perp)e^{{\rm i}k^+x^- -{\rm i}
\frac{2 \pi n^{\perp}}{L} x_\perp}\right]\,.
\nonumber
\eea
From the field \mbox{(anti-)}commutators one finds
\begin{equation}
\left[a_{ij}(p^+,n_\perp),a^\dagger_{lk}(q^+,m_\perp)\right]=
\left\{b_{ij}(p^+,n_\perp),b^\dagger_{lk}(q^+,m_\perp)\right\}=
\delta(p^+ -q^+)\delta_{n_\perp,m_\perp}\delta_{il}\delta_{jk}\,.
\end{equation}
Notice that the compactification in $x^\perp$ means that
the transverse momentum modes are summed over a discrete
set of values $2\pi n^\perp/L$.  In order to have a finite 
matrix representation for the eigenvalue problem, we must 
truncate these sums at some fixed integers $\pm T$.  The
value of $T$ defines a physical transverse cutoff
$\Lambda_\perp=2\pi T/L$; however, given this definition,
$T$ can also be viewed as a measure of transverse resolution
at fixed $\Lambda_\perp$.  

The supercharges take the following forms:
\bea\label{TruncSch}
&&Q^+={\rm i}2^{1/4}\sum_{n^{\perp}\in {\bf Z}}\int_0^\infty dk\sqrt{k}\left[
b_{ij}^\dagger(k,n^\perp) a_{ij}(k,n^\perp)-
a_{ij}^\dagger(k,n^\perp) b_{ij}(k,n^\perp)\right]\,,\\
\label{Qminus}
&&Q^-=\frac{2^{7/4}\pi {\rm i}}{L}\sum_{n^{\perp}\in {\bf Z}}\int_0^\infty dk
\frac{n^{\perp}}{\sqrt{k}}\left[
a_{ij}^\dagger(k,n^\perp) b_{ij}(k,n^\perp)-
b_{ij}^\dagger(k,n^\perp) a_{ij}(k,n^\perp)\right]+\nonumber\\
&&+ {{\rm i} 2^{-1/4} {g} \over \sqrt{L\pi}}
\sum_{n^{\perp}_{i} \in {\bf Z}} \int_0^\infty dk_1dk_2dk_3
\delta(k_1+k_2-k_3) \delta_{n^\perp_1+n^\perp_2,n^\perp_3}\nonumber\\
&&\times\left\{
     {1 \over 2\sqrt{k_1 k_2}} {k_2-k_1 \over k_3}
[a_{ik}^\dagger(k_1,n^\perp_1) a_{kj}^\dagger(k_2,n^\perp_2)
b_{ij}(k_3,n^\perp_3)
-b_{ij}^\dagger(k_3,n^\perp_3)a_{ik}(k_1,n^\perp_1)
a_{kj}(k_2,n^\perp_2) ]\right.\nonumber\\
&&+{1 \over 2\sqrt{k_1 k_3}} {k_1+k_3 \over k_2}
[a_{ik}^\dagger(k_3,n^\perp_3) a_{kj}(k_1,n^\perp_1) b_{ij}(k_2,n^\perp_2)
-a_{ik}^\dagger(k_1,n^\perp_1) b_{kj}^\dagger(k_2,n^\perp_2)
a_{ij}(k_3,n^\perp_3) ]\nonumber\\
&&+{1 \over 2\sqrt{k_2 k_3}} {k_2+k_3 \over k_1}
[b_{ik}^\dagger(k_1,n^\perp_1) a_{kj}^\dagger(k_2,n^\perp_2)
a_{ij}(k_3,n^\perp_3)
-a_{ij}^\dagger(k_3,n^\perp_3)b_{ik}(k_1) a_{kj}(k_2,n^\perp_2) ]\nonumber\\
&&+({ 1\over k_1}+{1 \over k_2}-{1\over k_3})
[b_{ik}^\dagger(k_1,n^\perp_1) b_{kj}^\dagger(k_2,n^\perp_2)
b_{ij}(k_3,n^\perp_3)
+b_{ij}^\dagger(k_3,n^\perp_3) b_{ik}(k_1,n^\perp_1) b_{kj}(k_2,n^\perp_2)]
         \left. \frac{}{}\right\}\,. \nonumber \\
\eea
All sums over the transverse momentum indices are truncated
at $\pm T$.  The symmetric truncation with respect to positive
and negative modes aids in retaining a reflection parity
symmetry in the states.

The remaining step of the (S)DLCQ procedure~\cite{sakai95,alp99,alp99b}
is discretization in the longitudinal direction.  This is equivalent to
the choice of periodic boundary conditions in $x^-$~\cite{yamawaki}.
For a review of ordinary DLCQ, see~\cite{bpp98}.  Where DLCQ
and SDLCQ differ is in the construction of the discrete $P^-$.
In DLCQ this is done directly, but in SDLCQ it is the supercharge
that is discretized, with $P^-$ constructed from the anticommutator
of $Q^-$ in the supersymmetry
algebra.  The difference between the two discretizations disappears in the
continuum limit.  The advantage of the SDLCQ approach is that
the spectrum obtained from its discrete $P^-$ is explicitly
supersymmetric at any numerical resolution, whereas the DLCQ 
spectrum becomes supersymmetric only in the continuum limit.
For additional discussion of this point, see Ref.~\cite{hhlpt99}.
The specifics of the longitudinal discretization are presented in the
following section on numerical methods.

The spectrum can be classified according to three commuting
$Z_2$ symmetries of $Q^-$.  As described in~\cite{hhlpt99},
they are transverse parity
\begin{equation}\label{defparity}
P: a_{ij}(k,n^\perp)\rightarrow -a_{ij}(k,-n^\perp)\,,\qquad
        b_{ij}(k,n^\perp)\rightarrow b_{ij}(k,-n^\perp)\,;
\end{equation}
the $T$ symmetry defined by Kutasov~\cite{kutasov93}, which
we call $S$ symmetry, to avoid confusion with the transverse
cutoff $T$,
\begin{equation}
\label{defz2}
S: a_{ij}(k,n^\perp)\rightarrow -a_{ji}(k,n^\perp)\,,\qquad
        b_{ij}(k,n^\perp)\rightarrow -b_{ji}(k,n^\perp)\,;
\end{equation}
and the product of these two symmetries, $R=PS$.  We diagonalize
the supercharge separately in the four sectors defined by the
four possible combinations of $P$ and $S$ eigenvalues.  
This significantly reduces the size of the individual matrices required
for a given level of resolution.  The $P$ symmetry is associated
with a double degeneracy of the massive states, that is in
addition to the usual boson/fermion degeneracy of supersymmetry.
A demonstration of this is given in~\cite{hhlpt99}.

\section{Numerical Methods}
\label{numerical}

We convert the mass eigenvalue problem $2P^+P^-|M\rangle = M^2 |M\rangle$ 
to a matrix eigenvalue problem by introducing a discrete $P^-$ in
a basis where $P^+$ and $P^\perp=0$ are diagonal.
As discussed in the previous section, this is done
in SDLCQ by first discretizing the supercharge $Q^-$
and then constructing $P^-$ from the square of the supercharge:
$P^- = (Q^-)^2/\sqrt{2}$.
We have already introduced a finite discretization
in the transverse direction, characterized by the
compactification scale $L$ and cutoff or resolution $T$.
To complete the discretization of the supercharge, we introduce
discrete longitudinal momenta $k^+$ as fractions $nP^+/K$ of the
total longitudinal momentum $P^+$.  Here $n<K$ and $K$ are positive
integers.  It is in the nature of light-cone coordinates that the 
longitudinal momenta can be chosen positive, and that $n$ and
the number of partons are bounded by $K$.  The integer $K$
determines the resolution of the longitudinal discretization and is known
in DLCQ as the harmonic resolution~\cite{pb85}.
The remaining integrals in $Q^-$ are approximated by
a trapezoidal form.  The continuum limit in the longitudinal direction
is then recovered by taking the limit $K \rightarrow \infty$. 

In constructing the discrete approximation we drop the
longitudinal zero-momentum mode.  For some discussion of
dynamical and constrained zero modes, see the review~\cite{bpp98} 
and previous work~\cite{alpt98,hhlpt99}.
Inclusion of these modes would be ideal, but the techniques
required to include them in a numerical calculation
have proved to be difficult to develop, particularly because
of nonlinearities.   For DLCQ calculations that can be 
compared with exact solutions, the exclusion of
zero modes does not affect the massive spectrum~\cite{bpp98}.
In scalar theories it has been known for some time that 
constrained zero modes can give rise to dynamical symmetry 
breaking~\cite{bpp98} and work continues on the role of zero modes and near 
zero modes in these theories~\cite{thorn}.  
It is possible that a careful treatment of the 
dynamical zero mode of the gauge field $A^+$
could give rise to dynamical breaking of 
supersymmetry of the type suggested in~\cite{witten}. 

Our earliest SDLCQ calculations~\cite{alp99b} were done using a code written in
Mathematica and performed on a PC. This code was rewritten in C++ for
the work presented in~\cite{hhlpt99} and has now been 
substantially revised to reach higher resolutions.
We are able to generate the Hamiltonian matrix
for $K=2$ through 7 and for values of $T$ up to $T=9$ at $K=4$, 
decreasing to $T=3$ at $K=7$.
The actual dimension 
of the Fock basis as a function of the transverse and longitudinal 
resolutions is given in Table~\ref{tab1}. If only one symmetry sector is used,
the dimension of the Hamiltonian matrix to be diagonalized 
is roughly eight times smaller.
The absolute limit for the new code, on a Linux workstation with
2GB of RAM, is approximately 32 million states.
\begin{table}
\caption{\label{tab1}
The size of the Fock basis as a function of the longitudinal
resolution $K$ and transverse cutoff $T$.} 
\begin{tabular}{c|rrrrrr}
$T$ & $K=4$&5&6&7&8&9\\
\hline
1& 150&768&4108&22544&131830&775104\\
2& 522&4142& 34834 & 305016& 2753162& 25431056\\
3& 1262&13632& 156270 & 1866304& 22972270& \\
4& 2498&34160& 496106 & 7505592& & \\
5& 4358&72128& 1268230 & & & \\
6& 6970&135408&  & & & \\
7& 10462&233344&  & & & \\
8& 14962&376752&  & & & \\
9& 20598&577920&  & & & \
\end{tabular}
\end{table}

We extract several of the lowest eigenstates of the Hamiltonian
matrix by applying the Lanczos algorithm~\cite{Lanczos}.  This
requires a filtering process to remove spurious states, including
copies of the very lowest states, which appear before all the desired
states converge.\footnote{The appearance of spurious copies is
understood~\cite{Lanczos} as a failure of orthogonality due to accumulation
of round-off errors.}  The approach that we take is patterned
after the work of Cullum and Willoughby~\cite{Cullum}.
After $n$ iterations of the Lanczos algorithm, we have an
$n\times n$ tridiagonal representation $A$ of the original
Hamiltonian matrix.  The spurious eigenvalues of $A$ appear
as degenerate copies or are found in the spectrum
of the matrix obtained from $A$ by removing the first row
and the first column.  These two possibilities are easily
checked because the tridiagonal matrices are easily diagonalized
in full, even though $n$ becomes large, on the order of several
thousand.  The remaining ``good'' eigenvalues are checked for convergence,
with the number of iterations extended until the desired
number or range of converged eigenvalues is obtained. 
The eigenvectors are readily obtained from the eigenvectors
of the tridiagonal matrix $A$.  The only difficulty with the 
filtering process is the possibility of a ``false negative,''
when a nearly degenerate eigenstate is incorrectly flagged
as spurious; we found this type of event to be rare, but, of
course, the frequency is dependent on the spectral density
and the specific filtering criteria.  

\begin{figure}
\centerline{\psfig{file=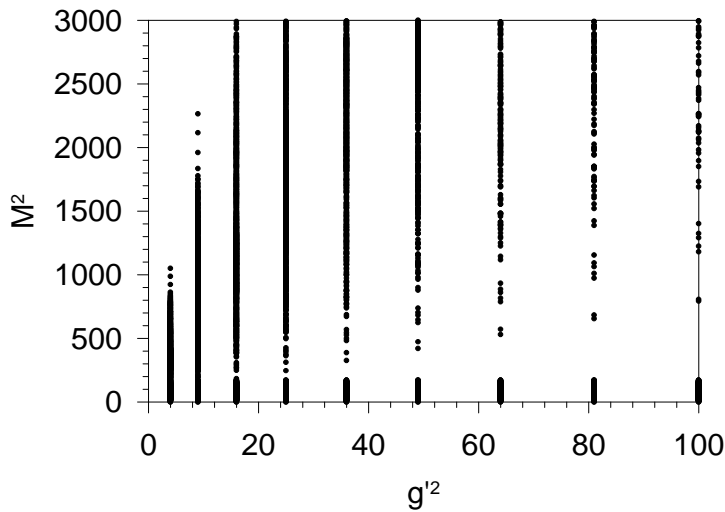,width=4in}}
\caption{The full spectrum of $M^2$ in units of $4 \pi^2 / L^2$ at
resolutions $K=5$ and $T=3$ in the $S=+1$, $P=+1$ symmetry sector. 
\label{full}}
\end{figure}

In Fig.~\ref{full} we show the full spectrum that we obtain as a function of a
dimensionless coupling $g'=g\sqrt{NL/4\pi^3}$.  This figure shows several 
striking features that we want to analyze in more detail.  First, we 
observe that the spectrum splits into two bands. At low mass the spectrum is
seen as a band of constant height in mass squared for all $g'$. The upper part 
of the spectrum appears as a band that grows in width and in mass squared. 
It is very clear that there is different physics at work in these two bands,
and we will analyze them separately. It is also important to note that 
different numerical methods are required to analyze these sectors. The 
lowest masses can be found with use of Lanczos diagonalization 
methods~\cite{Lanczos}; however, the lowest states in the upper band 
can only be found from a full diagonalization of the Hamiltonian.  

We refer to the lower band in Fig.~\ref{full} as the 
{\em intermediate-coupling
region}, for reasons that will become clear in the next section.  We
call the upper band the {\em strong-coupling region}, because these states
vary strongly with the coupling.
We study them in the strong-coupling limit in Sec.~\ref{scoupling}.

\section{Intermediate Coupling}
\label{Icoupling}

\subsection{Mass spectrum}

\begin{figure}[ht]
\begin{tabular}{cc}
\psfig{file=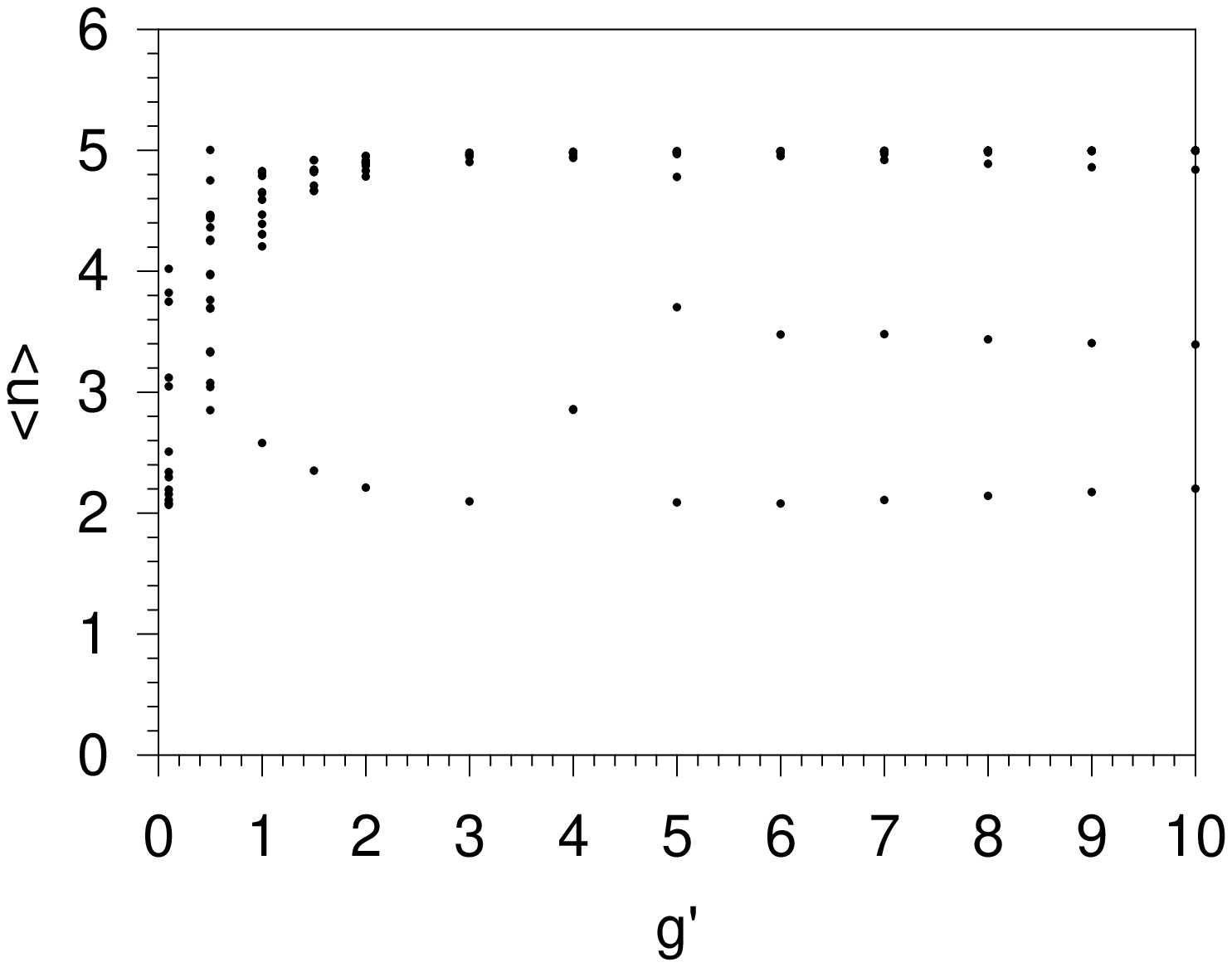,width=3.2in}  &
\psfig{file=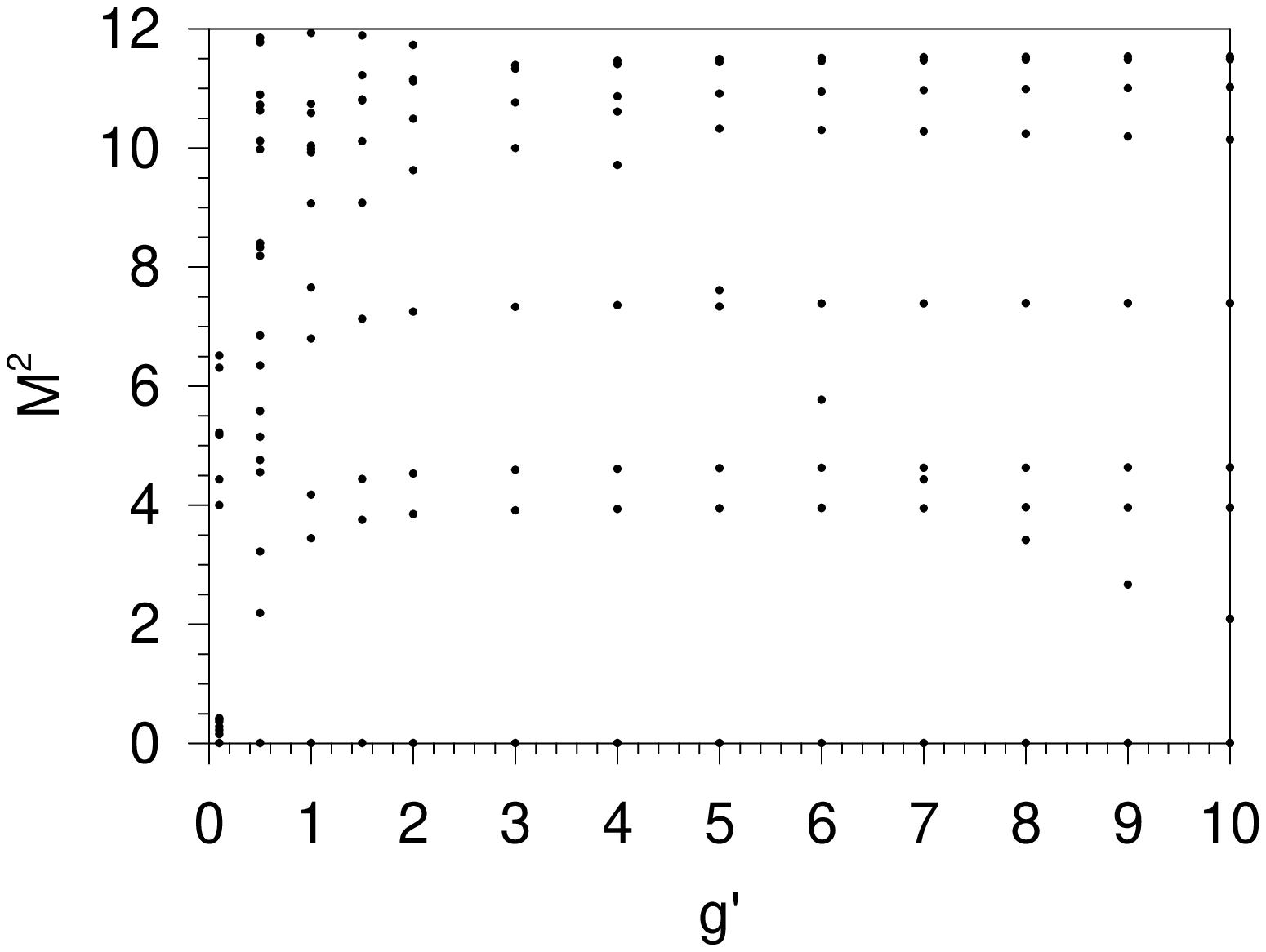,width=3.2in}  \\
(a) & (b)
\end{tabular}
\caption{Plots of (a) the average number of particles 
$\langle n \rangle$ and 
(b) bound-state mass squared $M^2$ in units of $4 \pi^2 / L^2$ as
functions of the coupling $g'$ at resolutions $K=5$ and $T=2$ 
in the $S=+1$, $P=-$ symmetry sector.
\label{low}}
\end{figure}

\begin{figure}[ht]
\begin{tabular}{cc}
\psfig{file=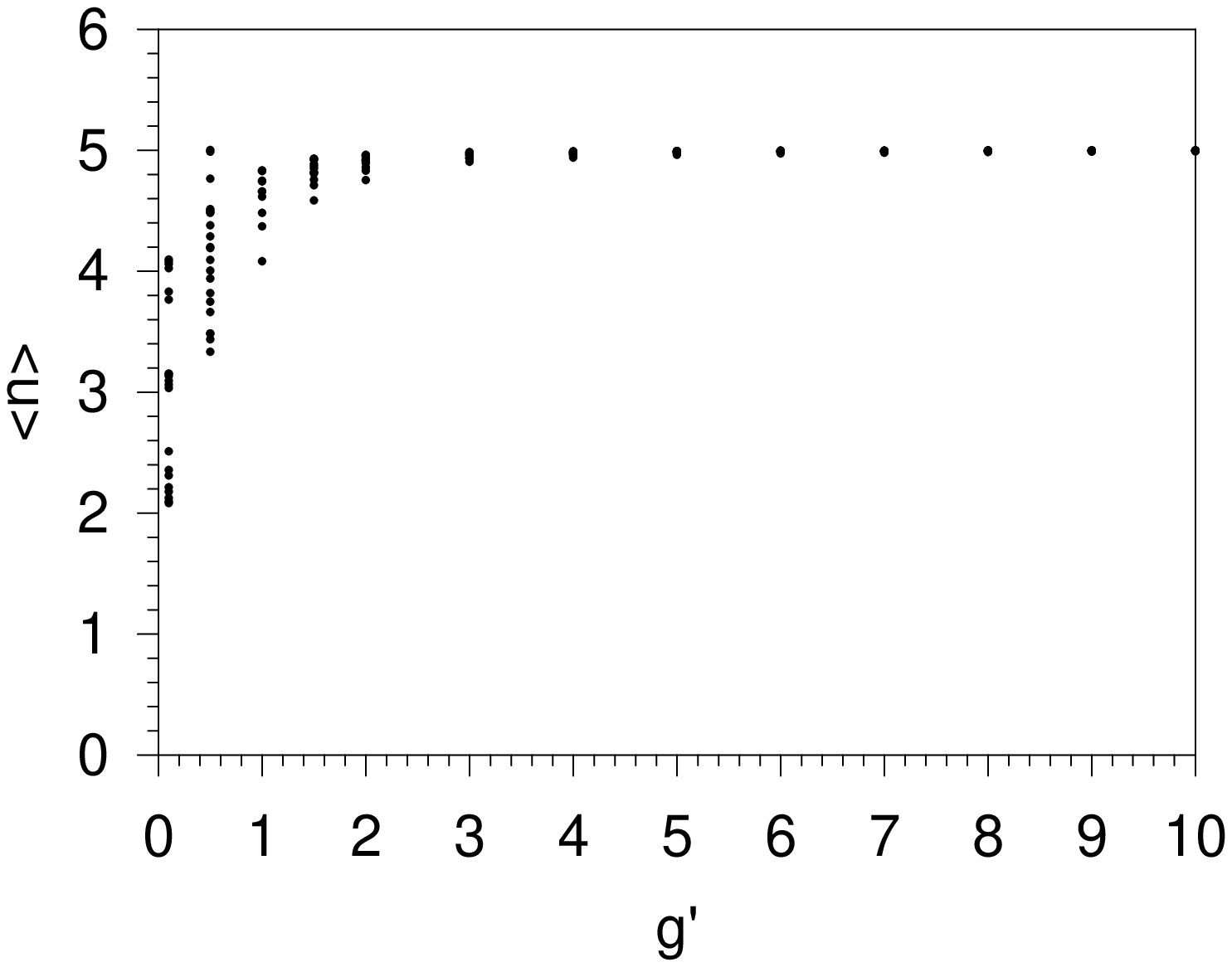,width=3.2in}  &
\psfig{file=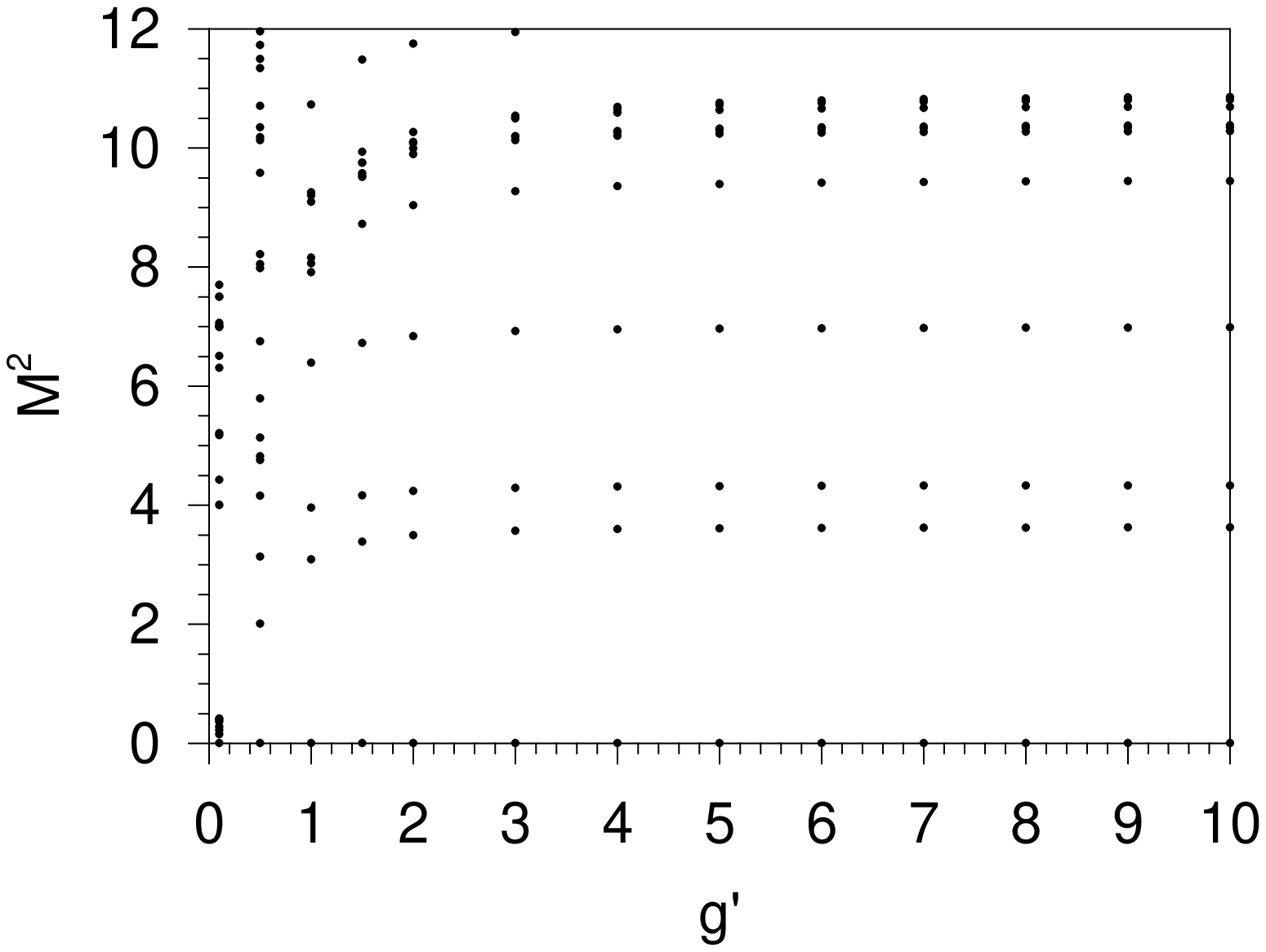,width=3.2in}  \\
(a) & (b)
\end{tabular}
\caption{Same as Fig.~\ref{low} but for $T=9$.
\label{high}}
\end{figure}

In Fig.~\ref{low} and Fig.~\ref{high} we show as functions of $g'$ the 
average number of partons and the mass squared of bound states in
the lower band of Fig.~\ref{full}.
We do this for low and high transverse resolutions. We see that in 
both cases the average number of particles grows very rapidly with $g'$,
and by $g'=1.5$ the average number is essentially equal to $K$. 
The resolution $K$ is the maximum number
of partons that are allowed in a state and corresponds to the situation 
where all the partons have one unit of longitudinal momentum.  
This should not come as a total surprise, since in the dimensionally reduced 
version of this model~\cite{sakai95,alp98a,alp98b} in 1+1 dimensions 
we already saw lighter and lighter states 
with more and more partons appearing as we went to higher resolution. 
This result suggests that already at intermediate 
couplings the states are saturating the SDLCQ approximation. That is, we are 
finding that, at every $K$, bound states have an average number of particles 
that is equal to $K$, the maximum number of particle allowed by the SDLCQ 
approximation. This implies that the actual states have an average 
number of particles that is significantly larger than the allowed maximum. 
Therefore, beyond intermediate values of the coupling, the SDLCQ approximation 
is missing a significant part of the wave function, and the calculation is 
becoming unreliable. For this reason we restrict our analysis in this
lower band to states with $g' = 0.5$ and 1.0\@. As we will see below,
the dependence of the spectrum on the longitudinal resolution
$K$ is stronger than what we have usually seen in SDLCQ, and this 
is because the approximation is losing significant parts of the 
wave function at these resolutions. 

The other effect that we see in Fig.~\ref{low} is a set of
states that we previously classified as unphysical~\cite{alp99b,hhlpt99}. 
Here these states appear at low masses at low transverse resolution, 
and the mass falls with increasing coupling.  We also see in Fig.~\ref{low} 
that there are states with low values of the average number of partons.
For these states we find $M^2 \propto \Lambda_\perp^2$, at least approximately.
We will find states in the strong-coupling sector that also grow rapidly with
transverse resolution as well, and we will return to these unphysical states 
when we discuss that sector.  In Fig.~\ref{high}, which shows 
results at a higher transverse resolutions, the states are no longer visible. 


A detailed analysis of a few intermediate-coupling bound states will 
be done in 
two steps. At fixed longitudinal resolution $K$, we will identify a sequence 
of states at different values of the longitudinal resolution $T$ that 
correspond to the same bound state. We make this identification using 
the properties of the state, most notably the mass, the average number of 
fermions, and the average momentum of the fermions. We have calculated a
large number of other average properties of the bound
states, but they are less useful in distinguishing states. We then 
plot this set of states as a function of $1/T$ and make a linear fit. 
The intercept is the mass at a particular value of $K$ with an 
infinite transverse-momentum cutoff. We then plot these 
cutoff-independent masses as a function of $1/K$ and make a second linear fit. 
The intercept is the mass at infinite longitudinal
resolution. By this method we identify the mass that is independent of the
transverse momentum cutoff and of the longitudinal resolution. 
These bound states are at a fixed value of the coupling, and the mass 
scale is set by the transverse length scale. 

As we discussed earlier, this theory has two exact symmetries,
parity $P$ and $S$ symmetry, and the problem of calculating the 
spectrum is divided into 4 sectors, $P=\pm 1$ and $S=\pm 1$. The $P$ 
symmetry gives rise to a doubling of the spectrum; 
this degeneracy is in addition to the fermion-boson degeneracy. 
These $P$-degenerate pairs are in fact not simply related. 
For example, in one sector a state may have an average number
of fermions $\langle n_F\rangle$ equal to 2, while in the 
other sector it might have $\langle n_F\rangle=0$. This property
proves  to be very useful in identifying sequences of bound states
at different $K$ and $T$. While two states might appear very similar in one 
$P$ sector, they can be very different in the other sector. 

The two $S$ sectors give rise to different sets of bound states, one for $S=+1$ 
and another for $S=-1$. For each bound-state mass there will be
two states with different properties, one corresponding to $P=+1$ and one
corresponding to $P=-1$.  We will present here the four lowest states in each 
$S$ sector. We label these states as $2.5\pm$, $3.0\pm$, $3.5\pm$ and 
$4.0\pm$ for convenience. 

\begin{figure}
\begin{tabular}{cc}
\psfig{file=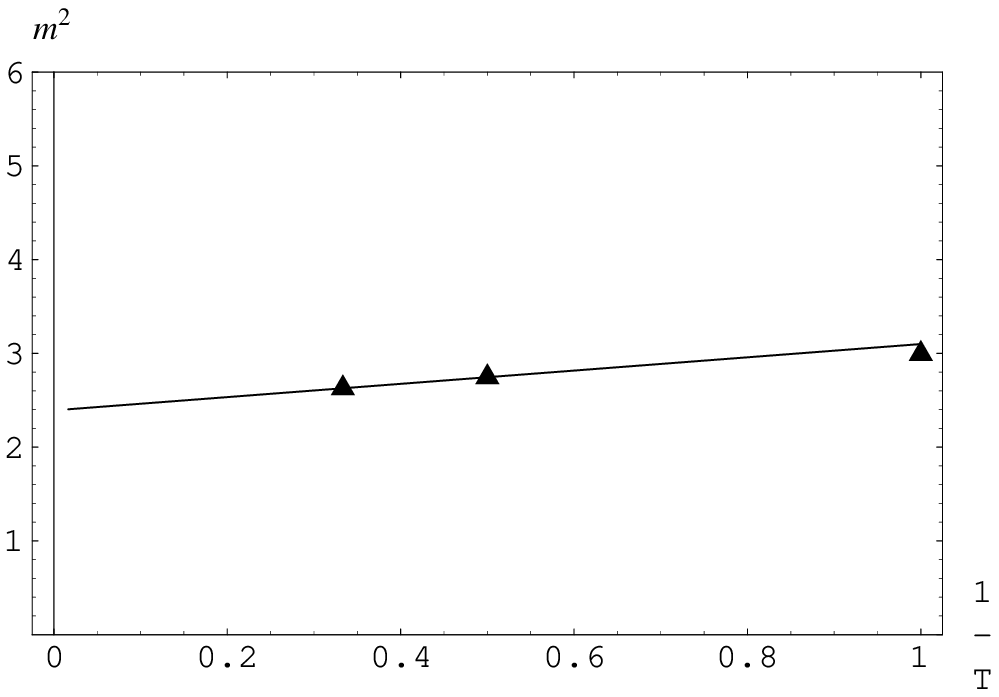,height=1.75in}  &
\psfig{file=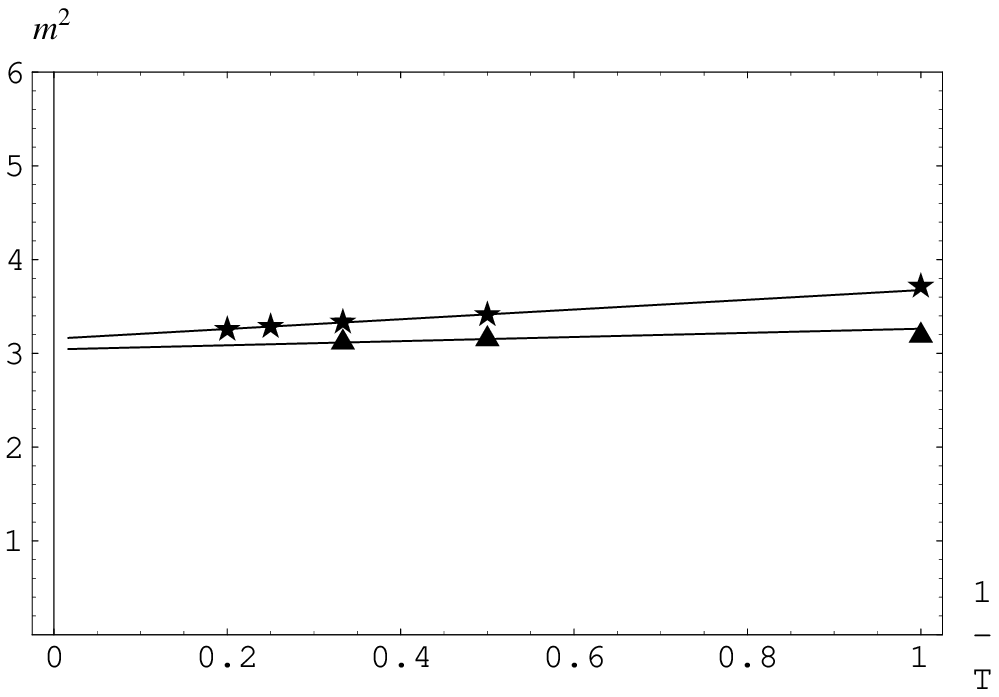,height=1.75in}  \\
(a) & (b)\\
\psfig{file=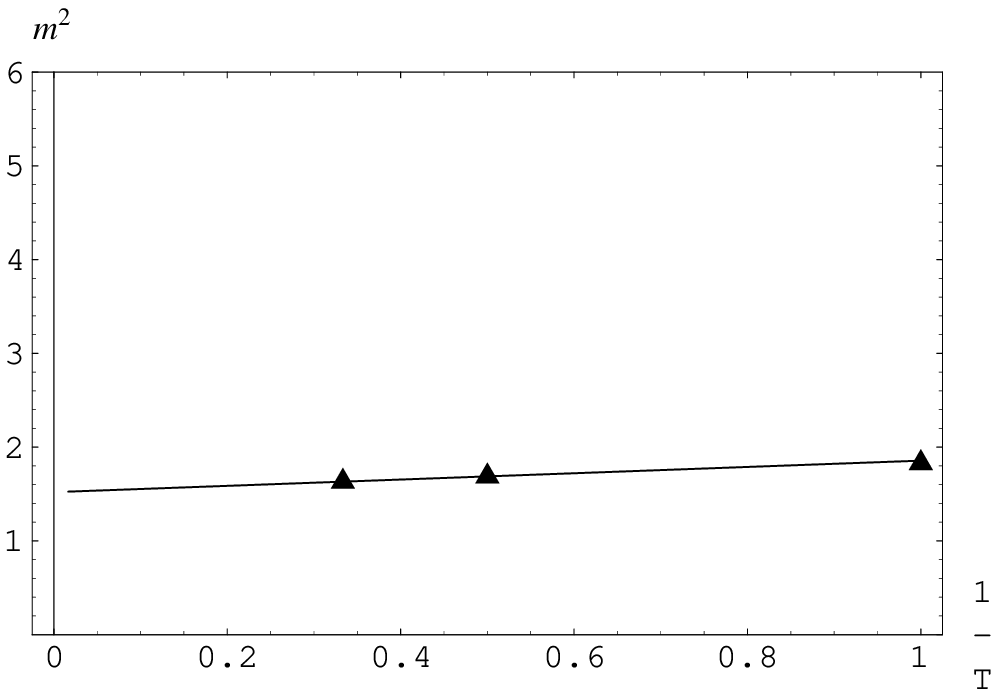,height=1.75in}  &
\psfig{file=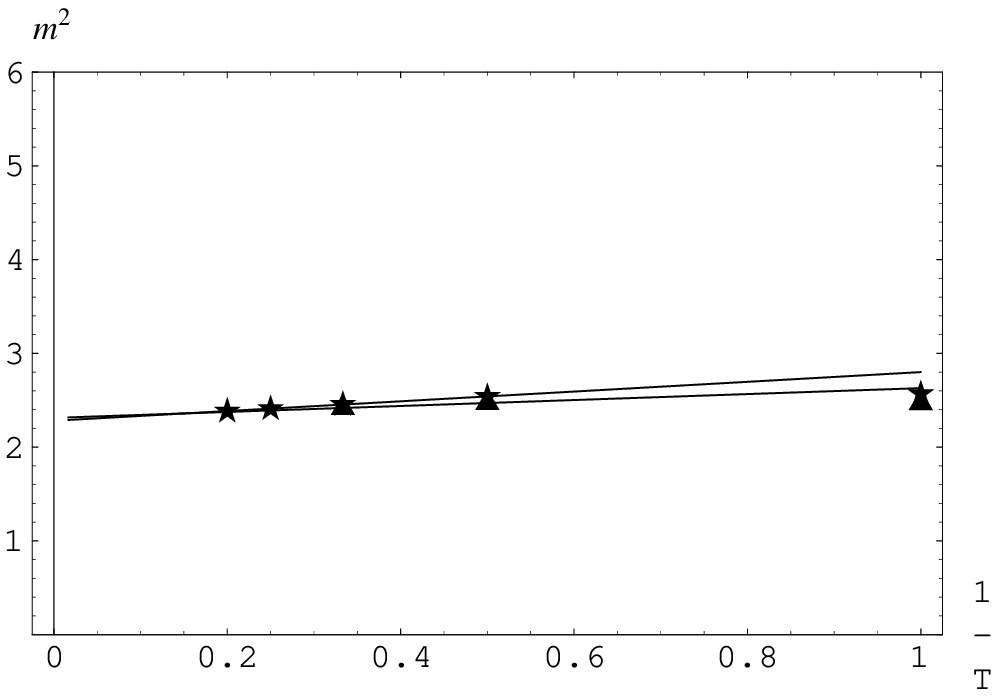,height=1.75in} \\
(c) & (d)\\
\psfig{file=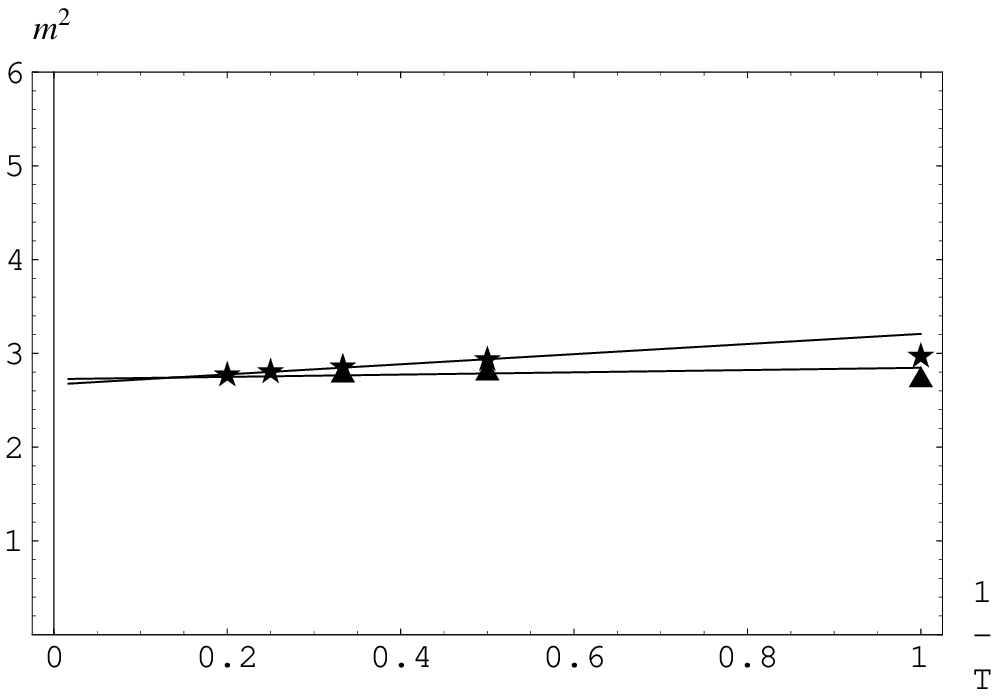,height=1.75in} &
\psfig{file=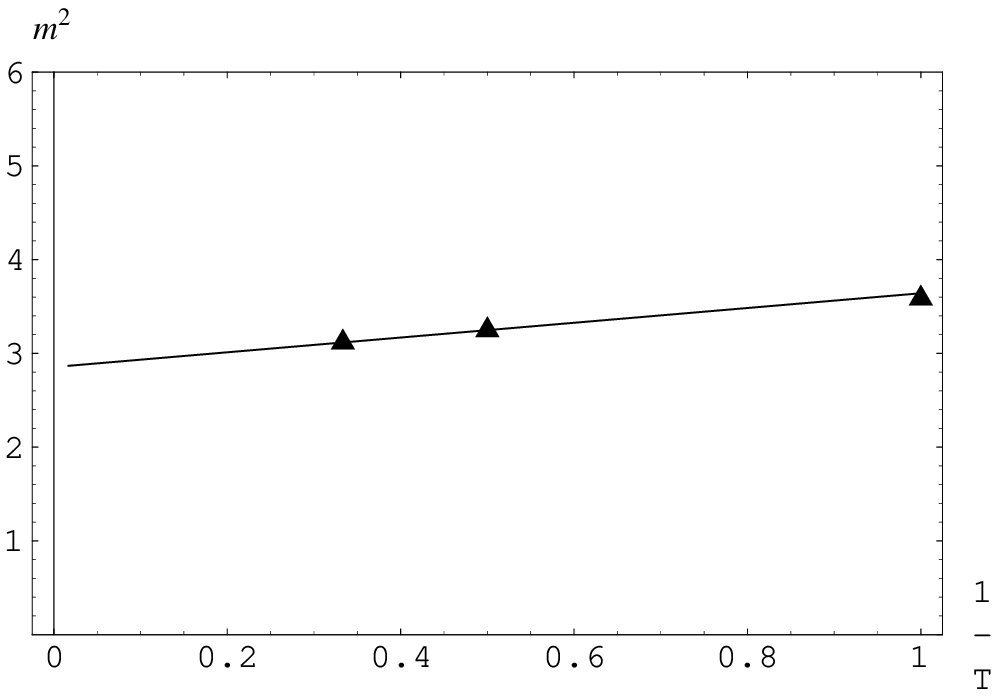,height=1.75in}  \\
(e) & (f)\\
\psfig{file=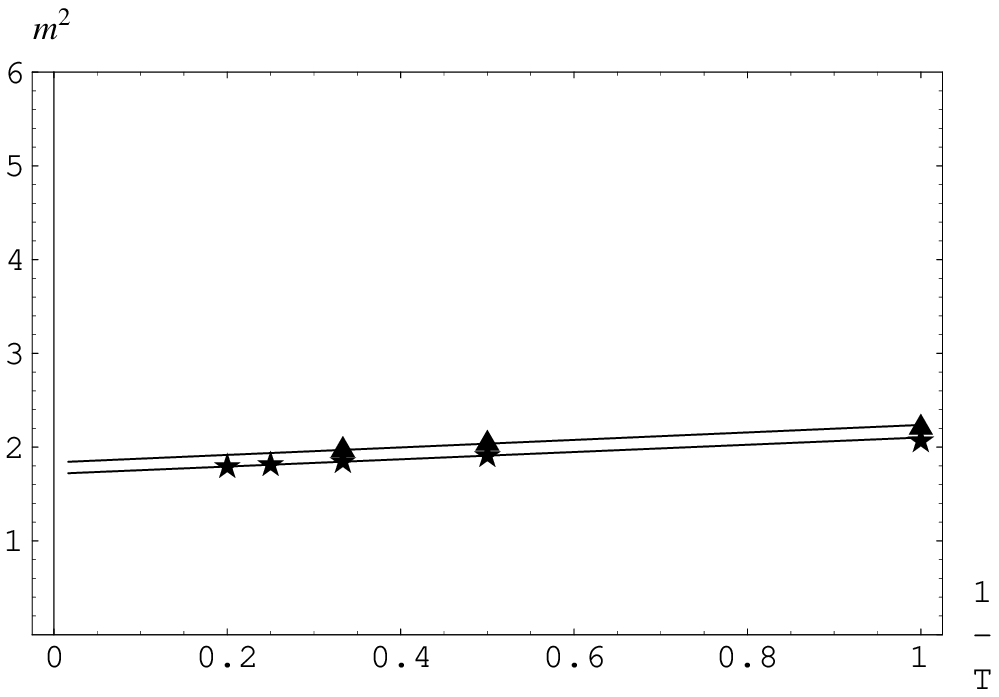,height=1.75in} &
\psfig{file=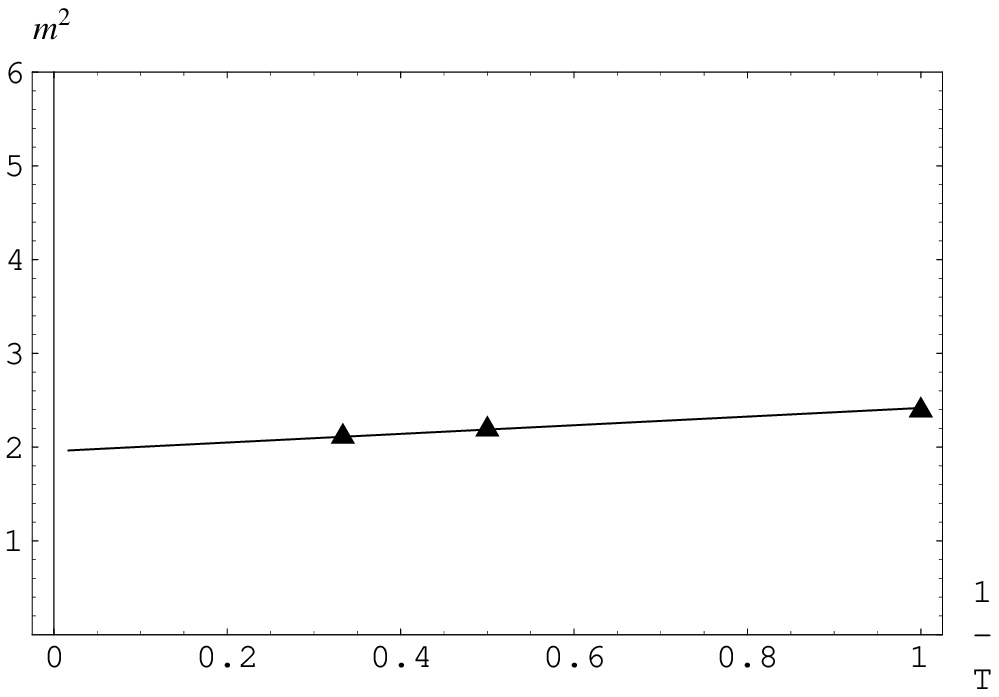,height=1.75in}  \\
(g) & (h)
\end{tabular}
\caption{Bound-state masses squared $M^2$ 
in units of $4 \pi^2 / L^2$  as functions of $1/T$ for
(a) states $2.5+$, $g'=1.0$;  (b) states $3.0+$, $g'=1.0$;
(c) states $2.5+$, $g'=0.5$;  (d) states $3.0+$, $g'=0.5$;
(e) states $2.5-$, $g'=1.0$;  (f) states $3.0-$, $g'=1.0$;
(g) states $2.5-$, $g'=0.5$;  (h) states $3.0-$, $g'=0.5$.
The longitudinal resolutions are $K=6$ (stars) and 7 (triangles).
\label{mass2.5-3.0}}
\end{figure}

As we mentioned earlier, in this theory the average number of
particles in a bound state grows as the mass of the bound state decreases. 
Therefore, the lightest massive bound state has a large number of particles. 
The maximum number of particles allowed by the approximation is the 
resolution, $K$. At $K=7$ and $K=6$ we are only able to include
$T$ up to 3 and 5, respectively. The two lowest mass states in each 
sector are only seen for $K=6$ or 7,
and for these cases we do not attempt to make
a linear fit in $1/K$ but rather we just use the average $T$ intercept for the
mass.  The $1/T$ curves for these states, $2.5\pm$ and $3.0\pm$, are shown in
Fig.~\ref{mass2.5-3.0} for $g'=0.5$ and $1.0$, and the properties are
given in Table~\ref{properties}.

\begin{figure}[ht]
\begin{tabular}{cc}
\psfig{file=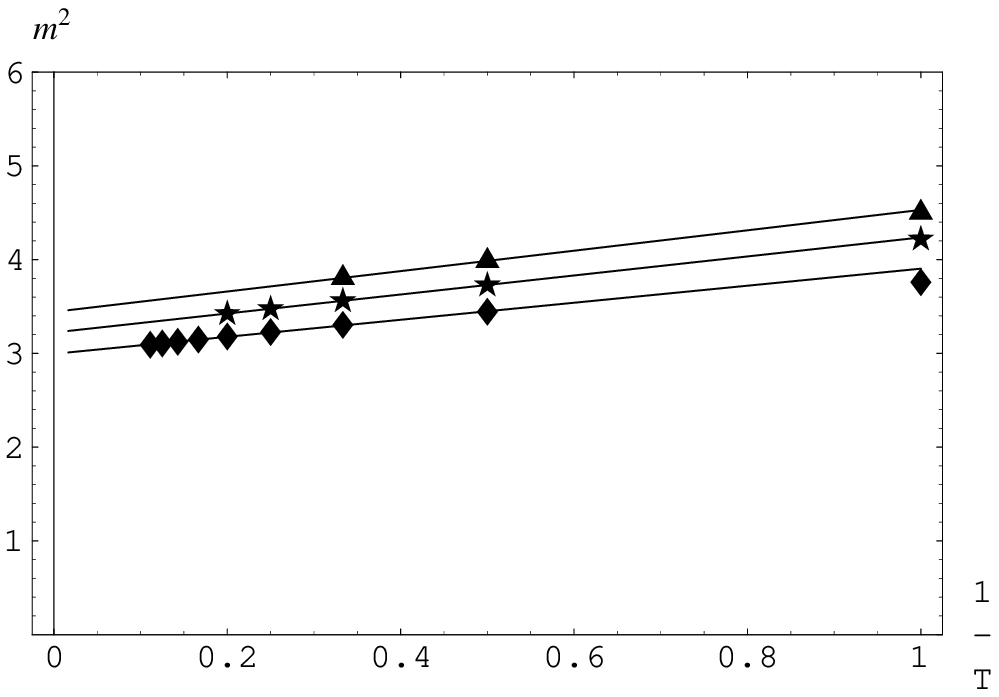,width=3.2in}  &
\psfig{file=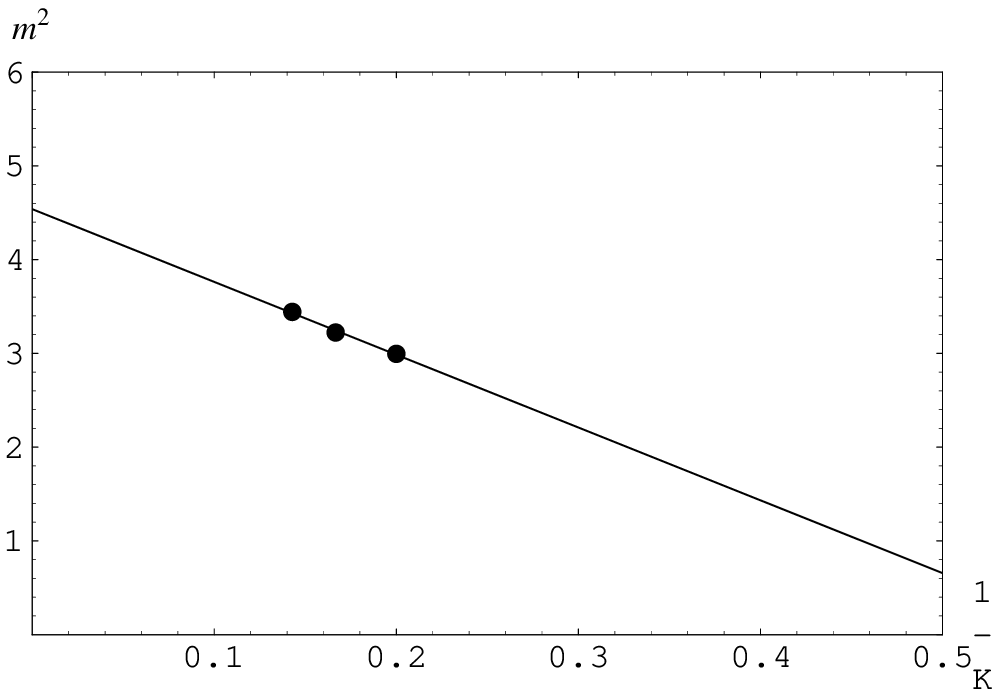,width=3.2in}  \\
(a) & (b)\\
\psfig{file=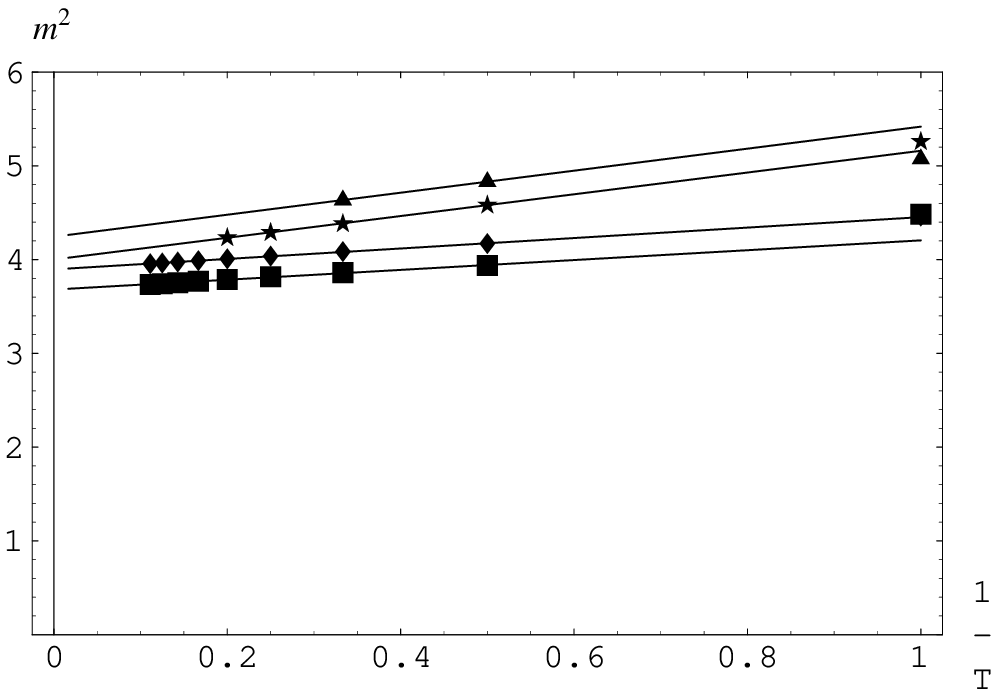,width=3.2in}  &
\psfig{file=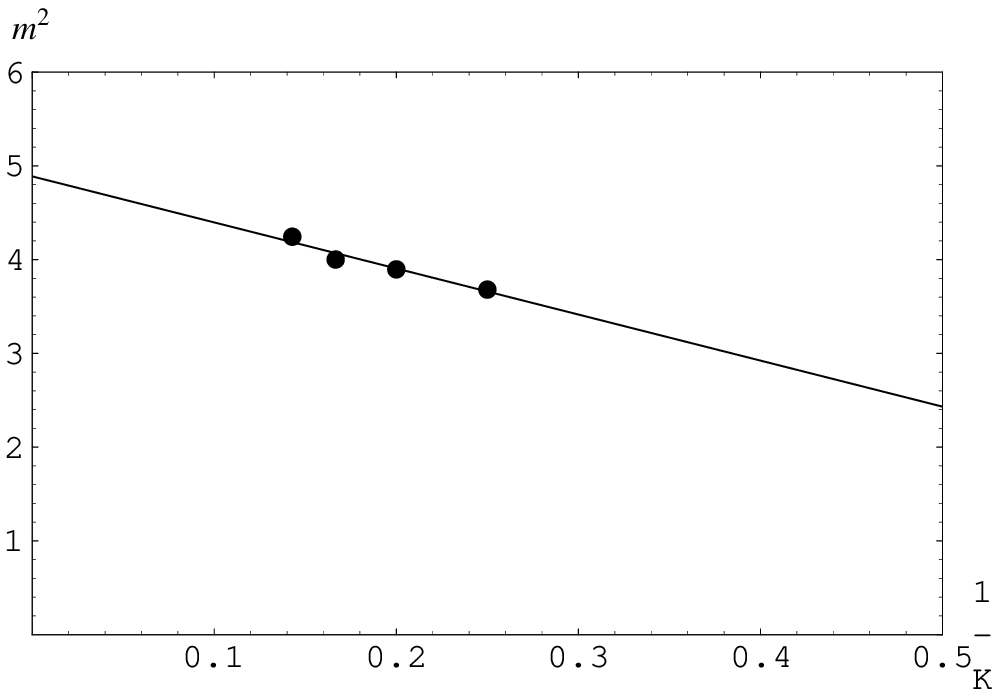,width=3.2in}  \\
(c) & (d)
\end{tabular}
\caption{Bound-state masses squared $M^2$ for the $S=+1$ sector 
in units of $4 \pi^2/ L^2$ at coupling $g'=1.0$  for 
(a) the state 3.5 as a function of $1/T$ for different values of $K$ 
and fit to straight lines,
(b) intercepts from (a) plotted as a function of $1/K$, 
(c) the state 4.0 as a function of $1/T$ for different values of $K$ 
and fit to straight lines,
and (d) intercepts from (c) plotted as a function of $1/K$.
The longitudinal resolutions are $K=4$ (squares), 5 (diamonds),
6 (stars), and 7 (triangles).
\label{g1S+}}
\end{figure}

\begin{figure}[ht]
\begin{tabular}{cc}
\psfig{file=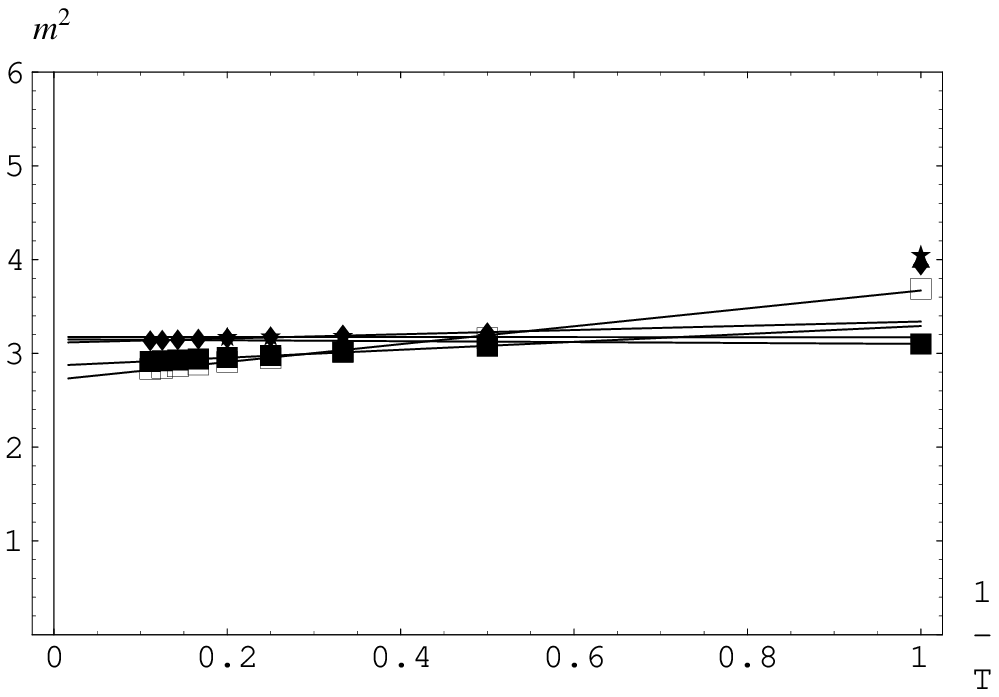,width=3.2in}  &
\psfig{file=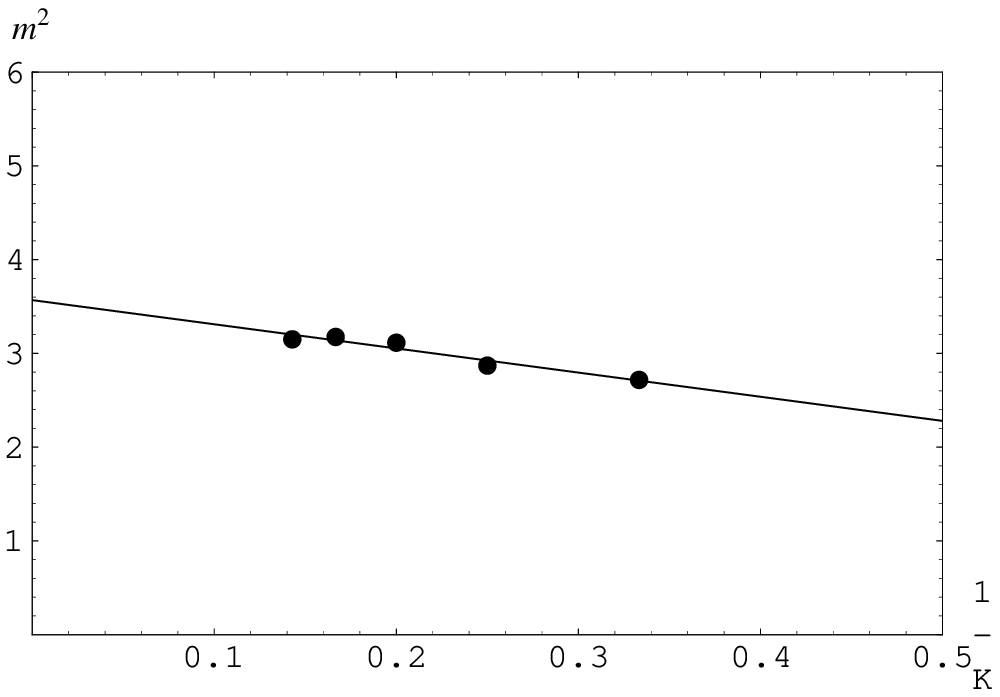,width=3.2in}  \\
(a) & (b)\\
\psfig{file=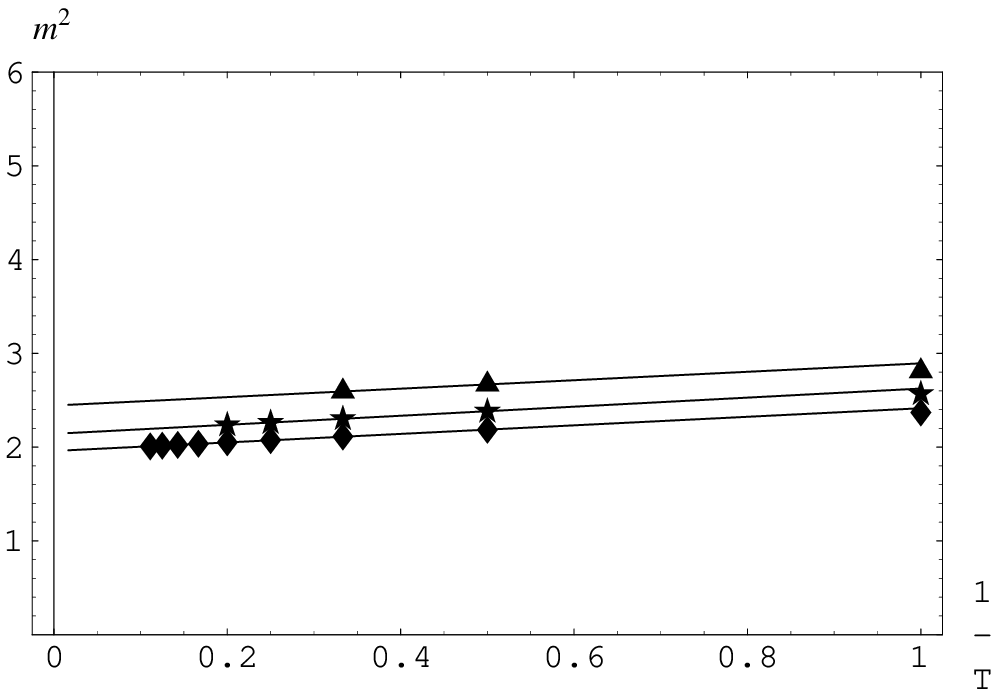,width=3.2in}  &
\psfig{file=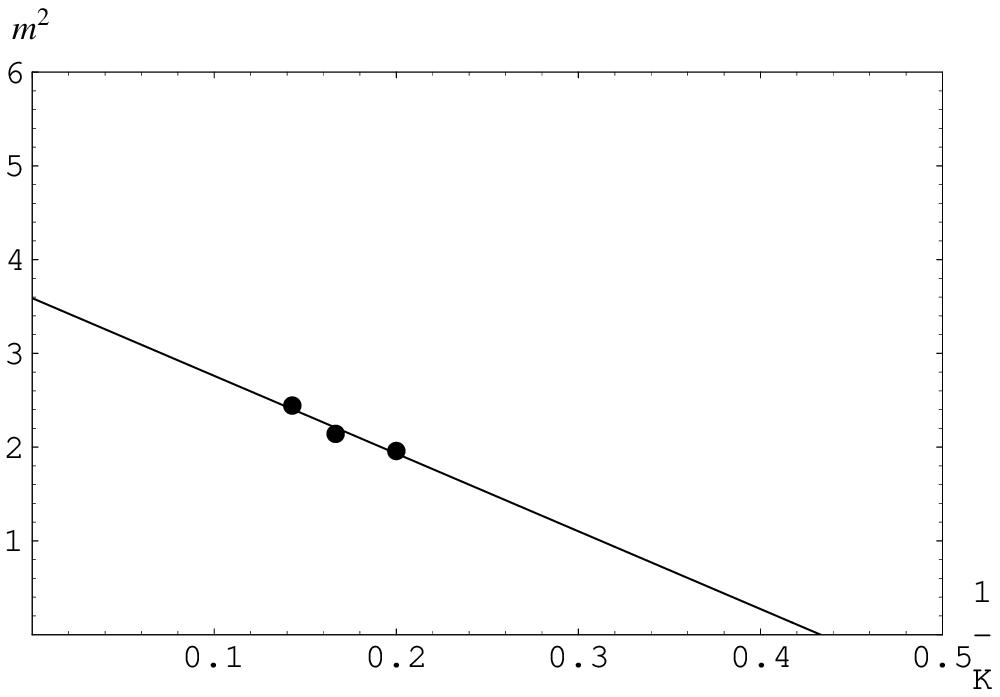,width=3.2in}  \\
(c) & (d)
\end{tabular}
\caption{Same as Fig.~\ref{g1S+} but for $g'=0.5$.
The open squares represent longitudinal resolution $K=3$.
\label{g.5S+}}
\end{figure}

\begin{figure}[ht]
\begin{tabular}{cc}
\psfig{file=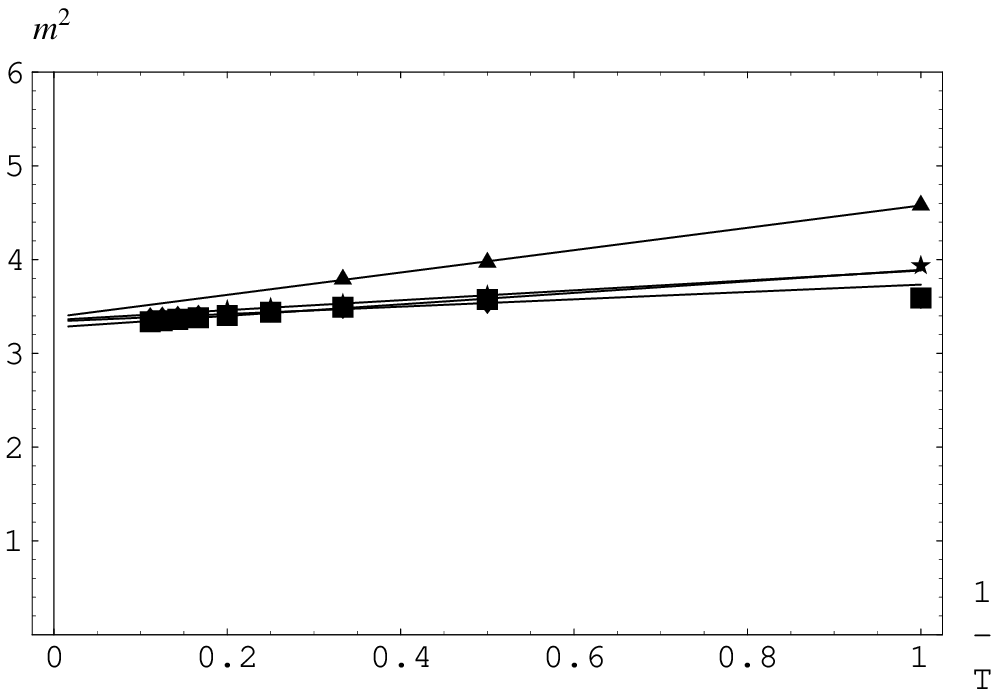,width=3.2in}  &
\psfig{file=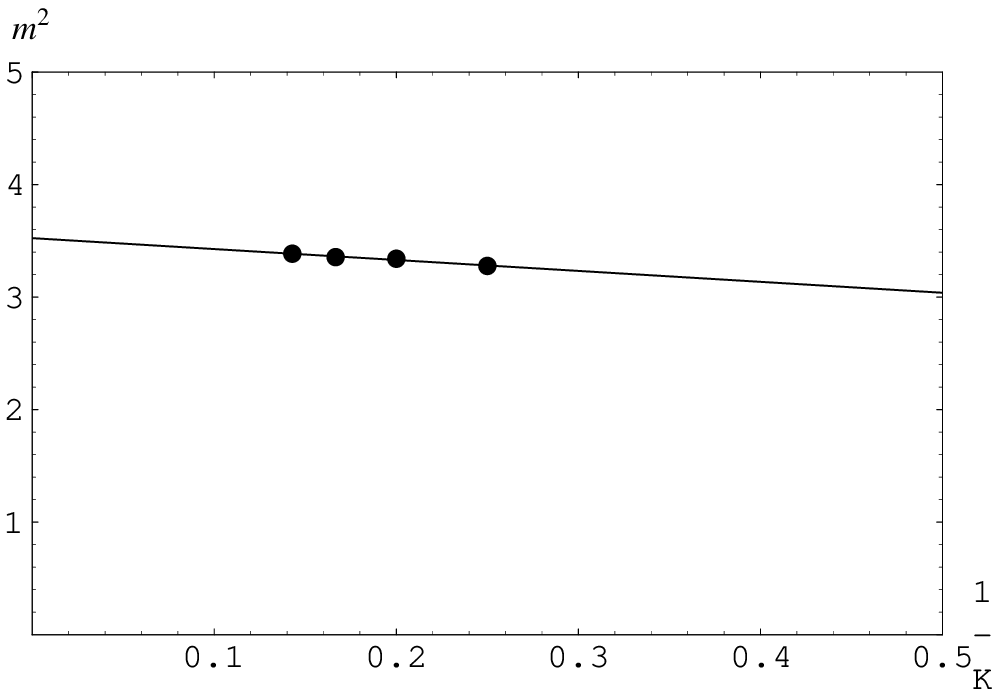,width=3.2in}  \\
(a) & (b)\\
\psfig{file=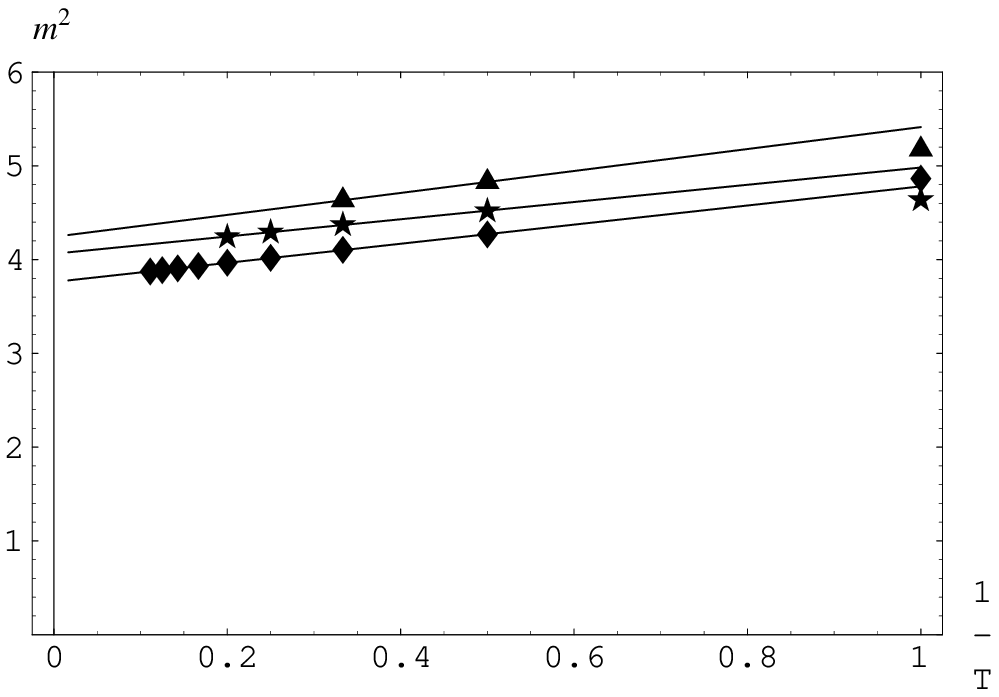,width=3.2in}  &
\psfig{file=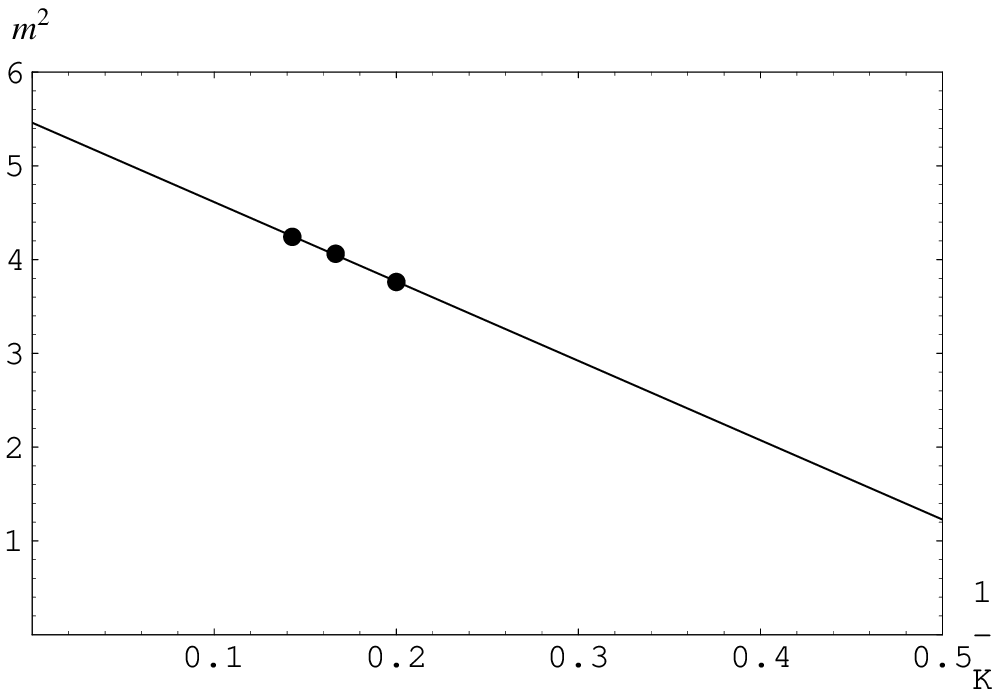,width=3.2in}  \\
(c) & (d)
\end{tabular}
\caption{Same as Fig.~\ref{g1S+} but for the $S-$ sector.
\label{g1S-}}
\end{figure}

\begin{figure}[ht]
\begin{tabular}{cc}
\psfig{file=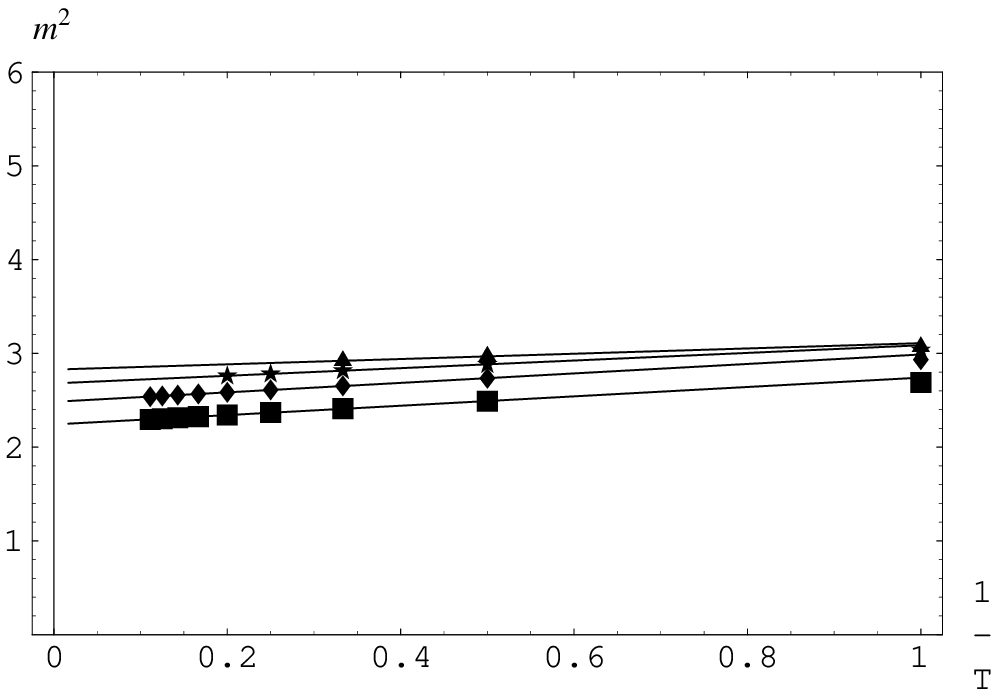,width=3.2in}  &
\psfig{file=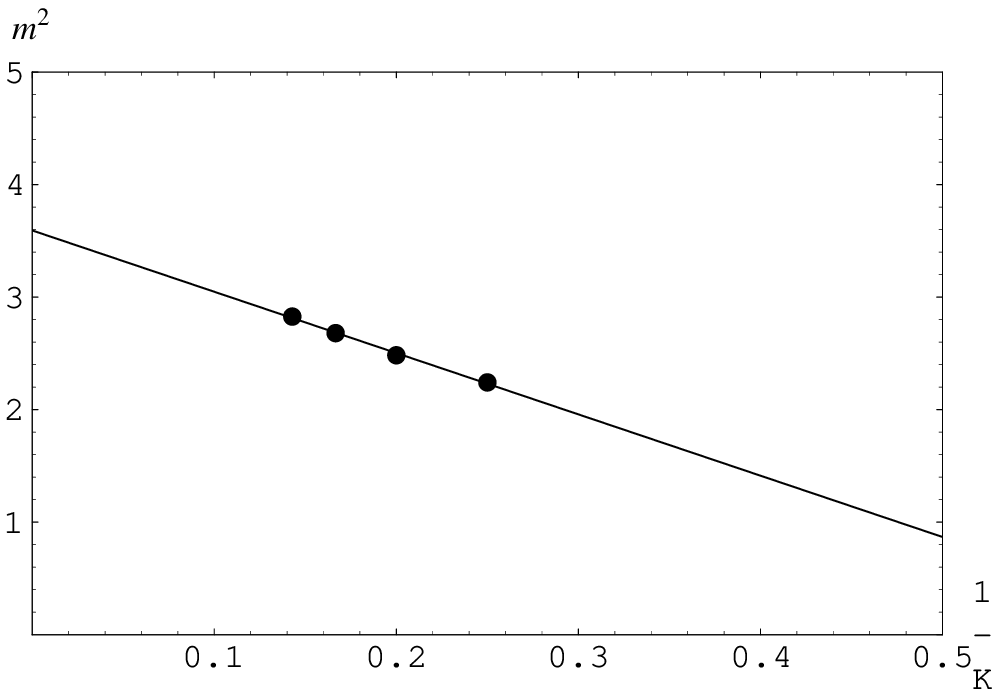,width=3.2in}  \\
(a) & (b)\\
\psfig{file=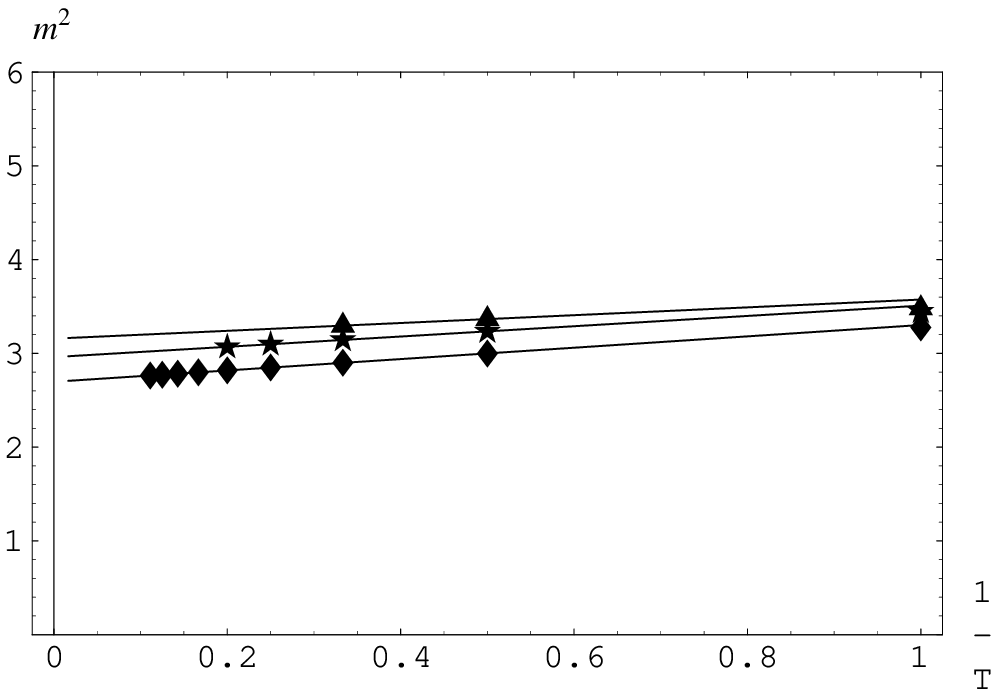,width=3.2in}  &
\psfig{file=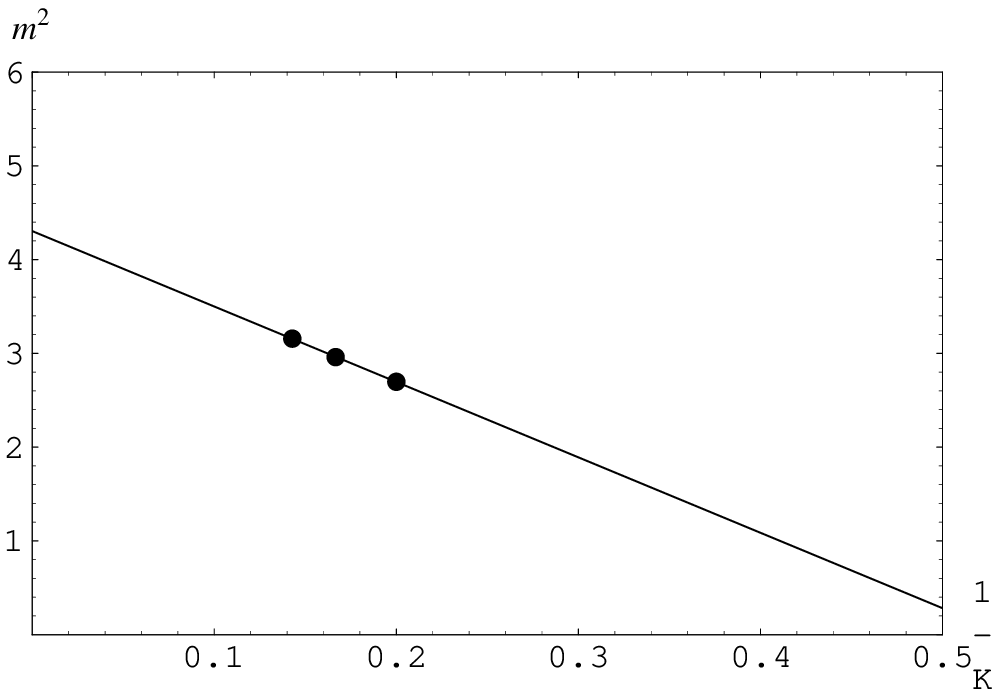,width=3.2in}  \\
(c) & (d)
\end{tabular}
\caption{Same as Fig.~\ref{g.5S+} but for the $S-$ sector.
\label{g.5S-}}
\end{figure}

Finally, we have states that can be identified at three values of $K$ and 
other states that can be identified at four values of $K$. In 
Fig.~\ref{g1S+} we show these states for the sector $S=+1$ and 
coupling $g'=1.0$, while in Fig.~\ref{g.5S+} we show the lowest
states at coupling $g'=0.5$. Similarly in Fig.~\ref{g1S-} and \ref{g.5S-} 
we show the lowest states in the $S=-1$ sector with $g'=1.0$ and $g'=0.5$,
respectively.  The properties of each are given in Table~\ref{properties2}.
\begin{table}[h]
\caption{Masses $M^2$ and average fermion number $\langle n_F\rangle$
for the states in Fig.~\ref{mass2.5-3.0}.  
The masses are in units of $4 \pi^2/ L^2$.
\label{properties}}
\begin{tabular}{c|c|c|c|c|c|c|c|c}
$M^2$&1.52&2.30&2.39&3.10&1.78&1.96&2.73&2.85 \\
\hline
$g'$&0.5&0.5&1.0&1.0&0.5&0.5&1.0&1.0  \\
\hline
$S$   &$+$&$+$&$+$&$+$&-&-&-&- \\
\hline
$\langle n_F\rangle$ with $P=+1$&2&0.5&2&0&2&2&2&2 \\
\hline
$\langle n_F\rangle$ with $P=-1$&2& 2\tablenotemark[1] &
     2\tablenotemark[1] & 2\tablenotemark[1] &2&0&2&0 \\
\end{tabular}
\tablenotetext[1]{These assignments are based on the symmetry
pattern observed in Table~\ref{properties2}, {\em i.e.},
low-mass, $S+$, $P-$ states appear to have two fermions.}
\end{table}

\begin{table}[h]
\caption{Masses $M^2$ and average fermion number $\langle n_F\rangle$
for the states in Figs.~\ref{g1S+}, \ref{g.5S+}, \ref{g1S-}, and \ref{g.5S-}.
The masses are in units of $4 \pi^2/ L^2$.
\label{properties2}}
\begin{tabular}{c|c|c|c|c|c|c|c|c}
$M^2$ &3.57&3.59&4.54&4.89 &3.59&4.30&3.52&5.46 \\
\hline
$g'$  &0.5&0.5&1.0&1.0 &0.5&0.5&1.0&1.0   \\
\hline
$S$   &$+$&$+$&$+$&$+$&$-$&$-$&$-$&$-$ \\
\hline
$\langle n_F\rangle$ with $P=+1$ &0&2&2&0 &2&2&2&2  \\
\hline
$\langle n_F\rangle$ with $P=-1$ &2&2&2&2 &2&0&2&0  \\
\end{tabular}
\end{table}


\subsection{Structure functions}

So far, we have only discussed the  integrated properties of the wave
functions of the individual states. It is, however, very instructive
to look at the structure functions. The fact that we saw very good 
convergence will be reflected in the invariance of the shape of the structure 
function of a given bound state seen at different $T$ and $K$. 

We use a standard definition of the structure functions
\begin{eqnarray}
\hat{g}_A(x,k^\perp)&=&\sum_q\int_0^1 dx_1\cdots dx_q \int_{-\infty}^{\infty}
dk^\perp_1\cdots dk^\perp_q 
\delta\left(\sum_{i=1}^q x_i-1\right)
\delta\left(\sum_{j=1}^q k^{\perp}_j\right)\nonumber \\
&&\qquad\times
\sum_{l=1}^q \delta(x_l-x)\delta(k^\perp_l-k^\perp)\delta^A_{A_l}
|\psi(x_1,k^\perp_1;\ldots x_q,k^\perp_q)|^2\,.
\end{eqnarray}
Here $A$ stands for either a boson or a fermion.
The sum runs over all parton numbers $q$, and the Kronecker delta 
$\delta^A_{A_l}$ selects partons with matching statistics $A_l$.
The discrete approximation $g_A$ to the structure function $\hat{g}_A$,
with resolutions $K$ and $T$ in longitudinal and transverse momentum, is
\begin{eqnarray}
{g}_A(n,n^\perp)&=&\sum_{q=2}^K\sum_{n_1,\ldots,n_q=1}^{K-q}
\sum_{n^\perp_1,\ldots,n^\perp_q=-T}^{T} 
\delta\left(\sum_{i=1}^q n_i-K\right)
\delta\left(\sum_{j=1}^q n^{\perp}_j\right)\nonumber \\
&&\qquad\times
\sum_{l=1}^q \delta^{n_l}_n\delta^{n^\perp_l}_{n^\perp}\delta^A_{A_l}
|\psi(n_1,n^\perp_1;\ldots n_q,n^\perp_q)|^2\,.
\end{eqnarray}
The functions $g_{A}(n,n^\perp)$ are normalized so that summation over both 
arguments gives the average boson or fermion number; 
their sum is then the average parton 
number, and we compute these sums as a test.

We focus on the state in the $S=-1$ 
sector which has a continuum mass of $M^2=3.52$. 
This state will have different 
manifestations at different cutoffs $K$ and $T$, and we have to ask
the following questions. Firstly, are the structure functions
stable enough against variations in both cutoffs to identify states at
different $K$ and $T$, and, secondly, are they distinct enough for different
states, so that we can distinguish between states that are close in mass
and other integrated properties of their wave function?

\begin{figure}
\begin{tabular}{cc}
\psfig{file=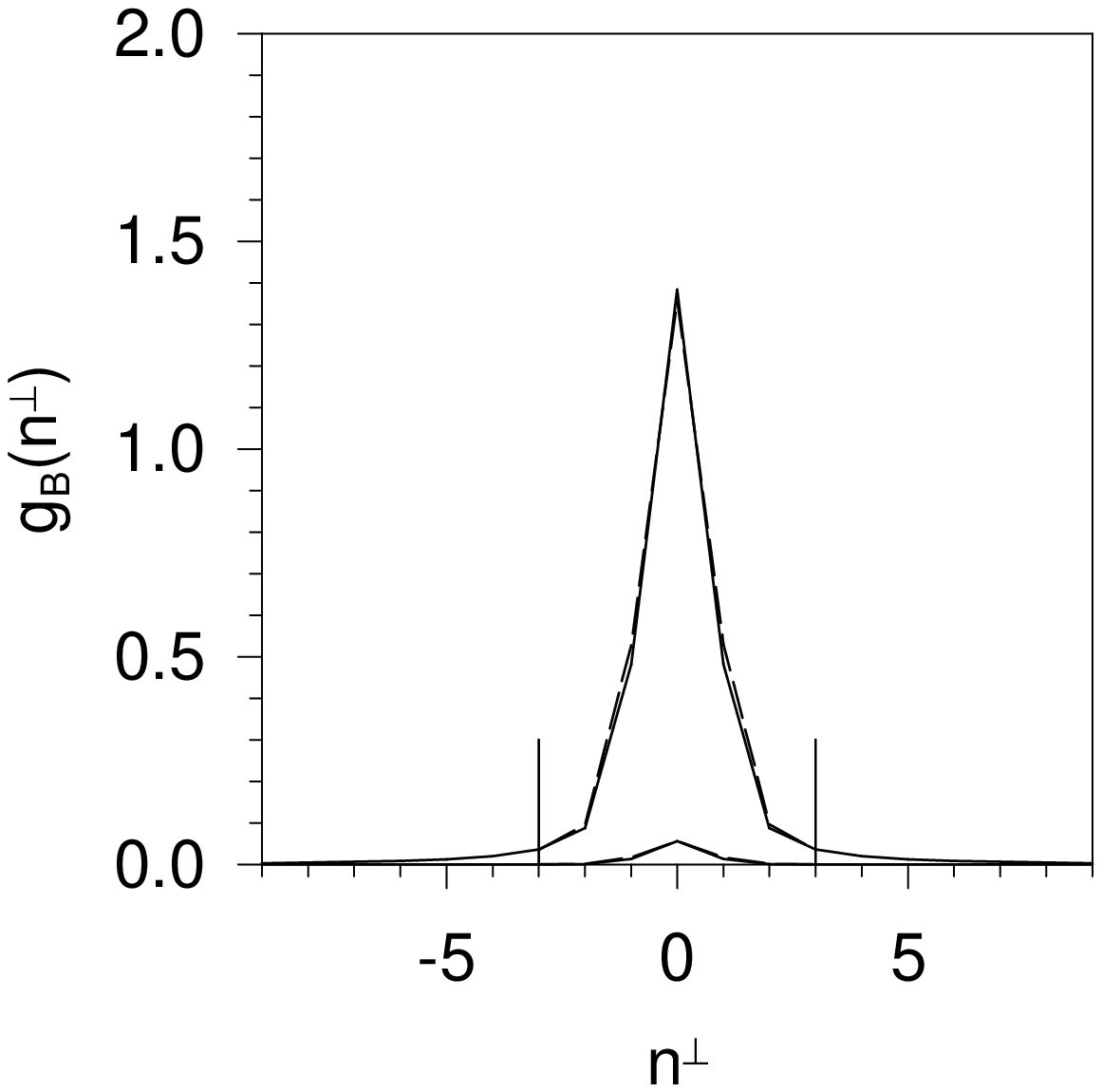,width=3.2in}  &
\psfig{file=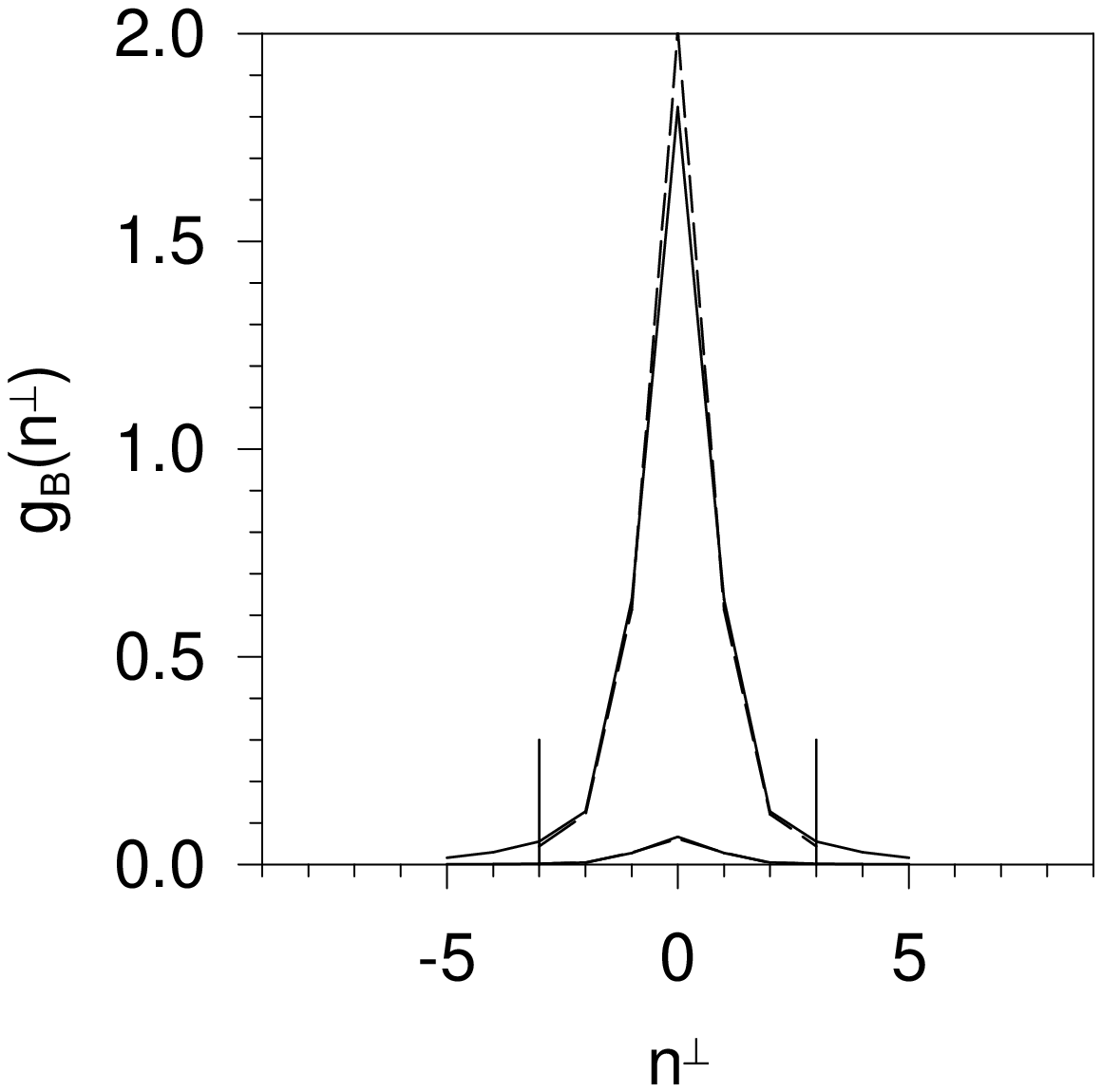,width=3.2in}  \\ 
(a) & (b)\\
\psfig{file=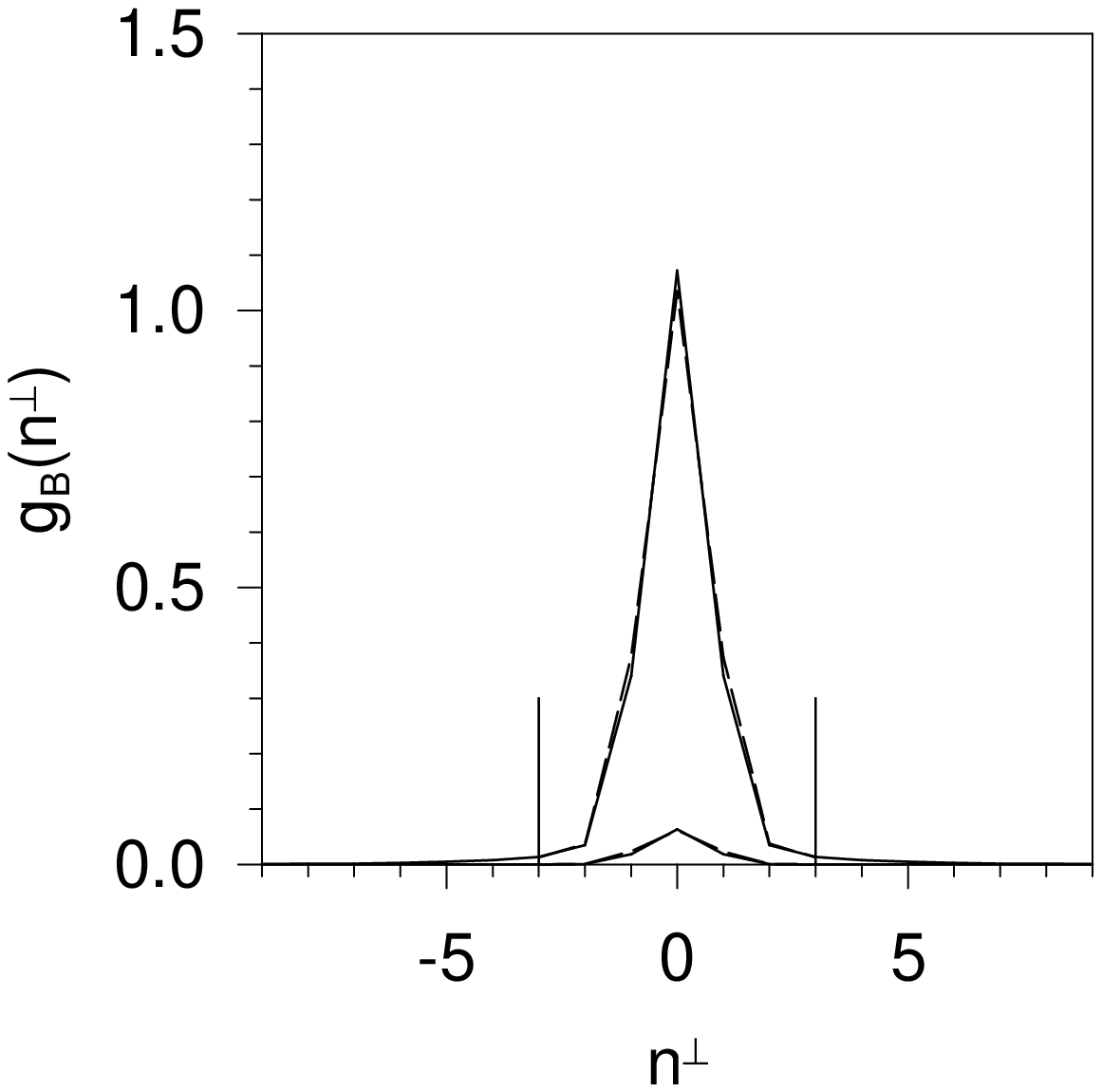,width=3.2in} & 
\psfig{file=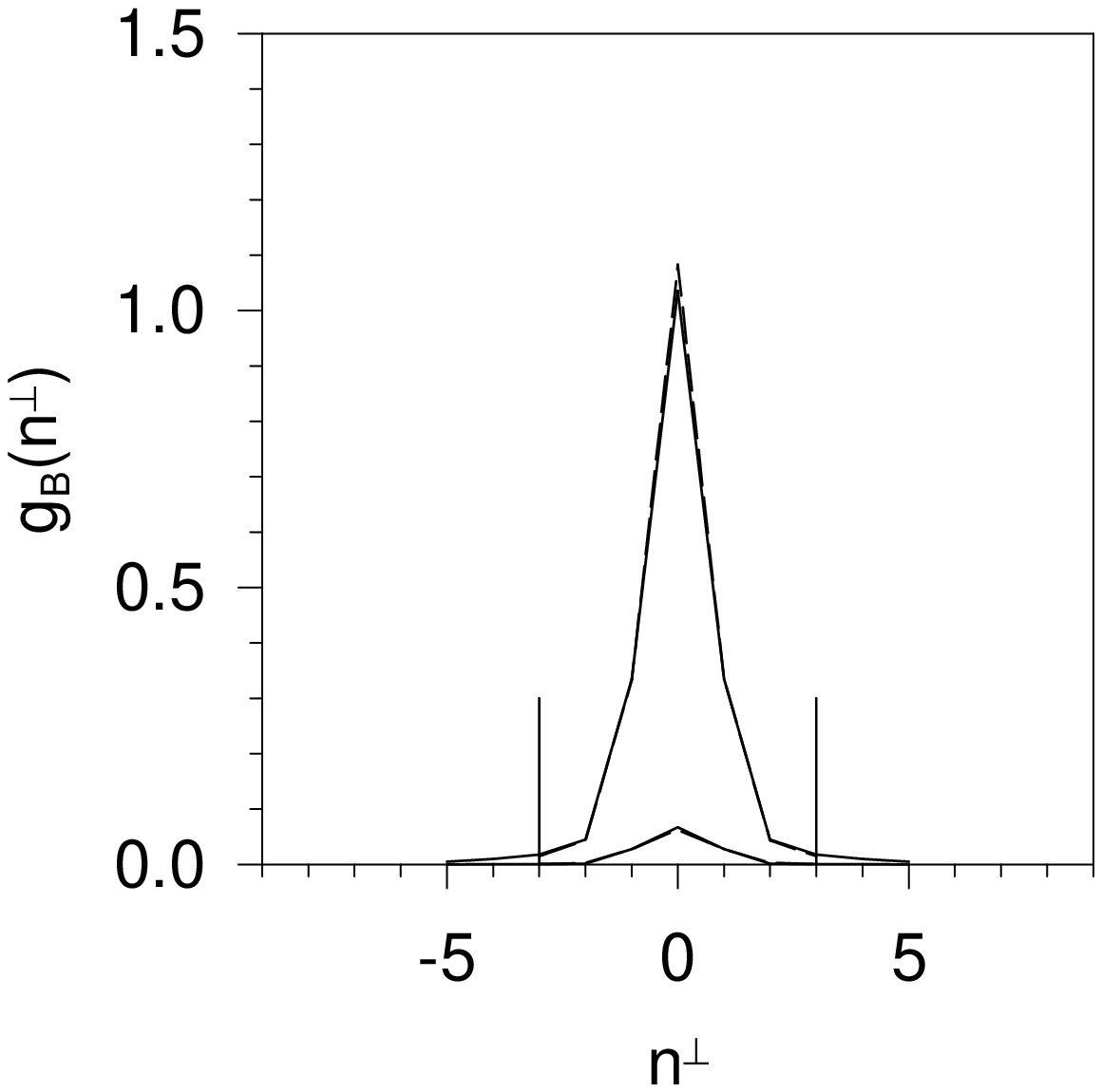,width=3.2in}  \\
(c) & (d)
\end{tabular}
\caption{Structure functions for the bound state with mass $M^2=3.52$
at different $K$ and $T$.
Top row: boson structure functions.
Left (a): $K=5$ at $T=3$ (dashed line) and $T=9$ (solid line).
Right (b): $K=6$ at $T=3$ (dashed line) and $T=5$ (solid line).
Bottom row: fermion structure functions.
Left (c):  $K=5$ at $T=3$ (dashed line) and $T=9$ (solid line).
Right (d):  $K=6$ at $T=3$ (dashed line) and $T=5$ (solid line).
The vertical lines mark the range in $n^\perp$ for T=3.
\label{SFs}}
\end{figure}

To address the first question, consider Fig.~\ref{SFs}, where 
we present plots of the boson and fermion structure functions of the state.
It is important to keep in mind that in each plot there are {\em four} 
curves, corresponding to different transverse cutoff values and to different
amounts of longitudinal momentum.
The curves with the higher amplitude in each plot 
are the probability to find the parton with a longitudinal momentum of 
$n=1$, and the low-amplitude curves are the probability 
to find the same parton with $n=2$.
We suppress curves with $n\ge 2$, because of their much smaller amplitudes.

The first thing to notice is that the shapes of all states
are nearly invariant under a change of the transverse cutoff $T$.
Unfortunately, this means that the (dashed and solid) lines are almost 
indistinguishable. 
The only thing that changes is the actual cutoff in $n^\perp$: the
curves have support only up to a maximum $n^\perp_{\rm max}=T$. 
This might be also hard to see in the plots, and we put vertical lines 
in each plot at the points $n^\perp=\pm 3$, where the support of the structure 
function at the smaller transverse cutoff ends.  This behavior is 
contrary to the one in the longitudinal direction, where a larger
$K$ is synonymous with a better resolution of the wave function.
Here, however, a change of $T$ clearly is a change of the cutoff 
$\Lambda_{\perp}=2\pi T/L$, where $L$ is the fixed transverse box size.

Having established this remarkable stability in $T$, let us now look at
how the structure function changes as $K$ grows. In Fig.~\ref{SFs}
we have to compare the functions in the two upper plots. 
We see that they are very close in shape, although the peak values change.
A discussion of the differences in peak value is in order. In Fig.~\ref{SFs}, 
we see that the peak values are the same for the fermion structure functions 
at different $K$, but are different for the boson structure functions.
The reason for this becomes clear if we look at the properties of this 
state: the average fermion number stays constant with $K$,
whereas the average parton number grows considerably. The latter fact is 
what we see in Figs.~\ref{low} and \ref{high}, and it also tells us that 
we mainly add partons with vanishing $n_\perp$, as we go higher in $K$.
A quick look at the 
fermion structure functions in the lower row of Fig.~\ref{SFs}
reveals a consistent behavior. Again, we see
no dependence on the transverse cutoff $T$. The curves are a perfect match 
at different $K$. 

\begin{figure}
\centerline{\psfig{file=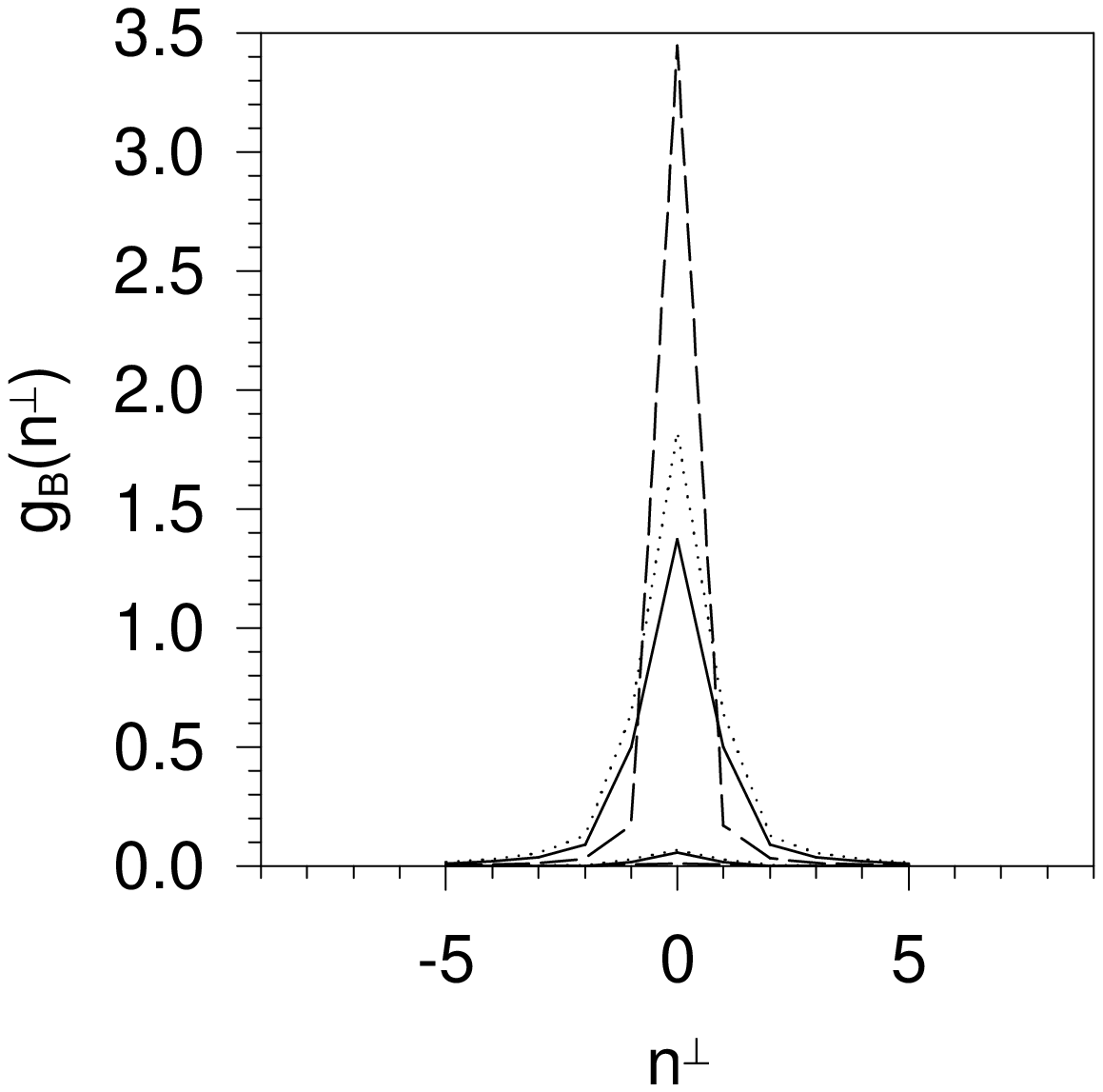,width=3.2in}}
\caption{Structure functions of two manifestations of the same state at
different $K=5,6$ and a distinct state at $K=6$:
the state with continuum mass $M^2=3.52$ for $K=5$ (solid line) and
$K=6$ (dotted line); 
the state with continuum mass $M^2=2.73$ for $K=6$ (dashed line).
For all cases the value of $T$ is 5, the symmetry sector
is $S=-1$, $P=+1$, and the coupling is $g'=1$.
\label{SFs2}}
\end{figure}

Now that the shape invariance of the structure functions has been extended
to changes in both $K$ and $T$, let us address the second question, of whether
the structure functions are different enough to distinguish states.
Consider Fig.~\ref{SFs2}, where the state with continuum mass squared
$M^2=3.52$ is plotted ({\em cf}.~Fig.~\ref{SFs}) at $T=5$ for $K=5$ and $K=6$, 
together with the next lightest state with $\langle n_F\rangle=2$ in both
parity sectors
at $K=6$, which has a mass $M^2=2.73$
according to Table~\ref{properties}. It is immediately clear 
that the latter state is different and 
cannot be a manifestation of the one with $M^2=3.52$: both the $n=1$ 
and the $n=2$ components of the structure function 
have different shapes, even if we would scale 
them to have the same peak values. The shape of the structure function of a 
state seems, therefore, a very characteristic and stable property of each 
individual state.

%
%

%
%

\section{Strong Coupling}
\label{scoupling}

Our analysis of the strong-coupling region (the upper band
of Fig.~\ref{full}) can only 
be considered preliminary at this stage, since these states are in the 
middle of the spectrum and one therefore needs a full diagonalization 
of the Hamiltonian to reach them. This severely limits the values
of $K$ and $T$ that we can consider. Nevertheless, we think it is worth 
taking a look at this part of the spectrum.

\begin{figure}[ht]
\begin{tabular}{cc}
\psfig{file=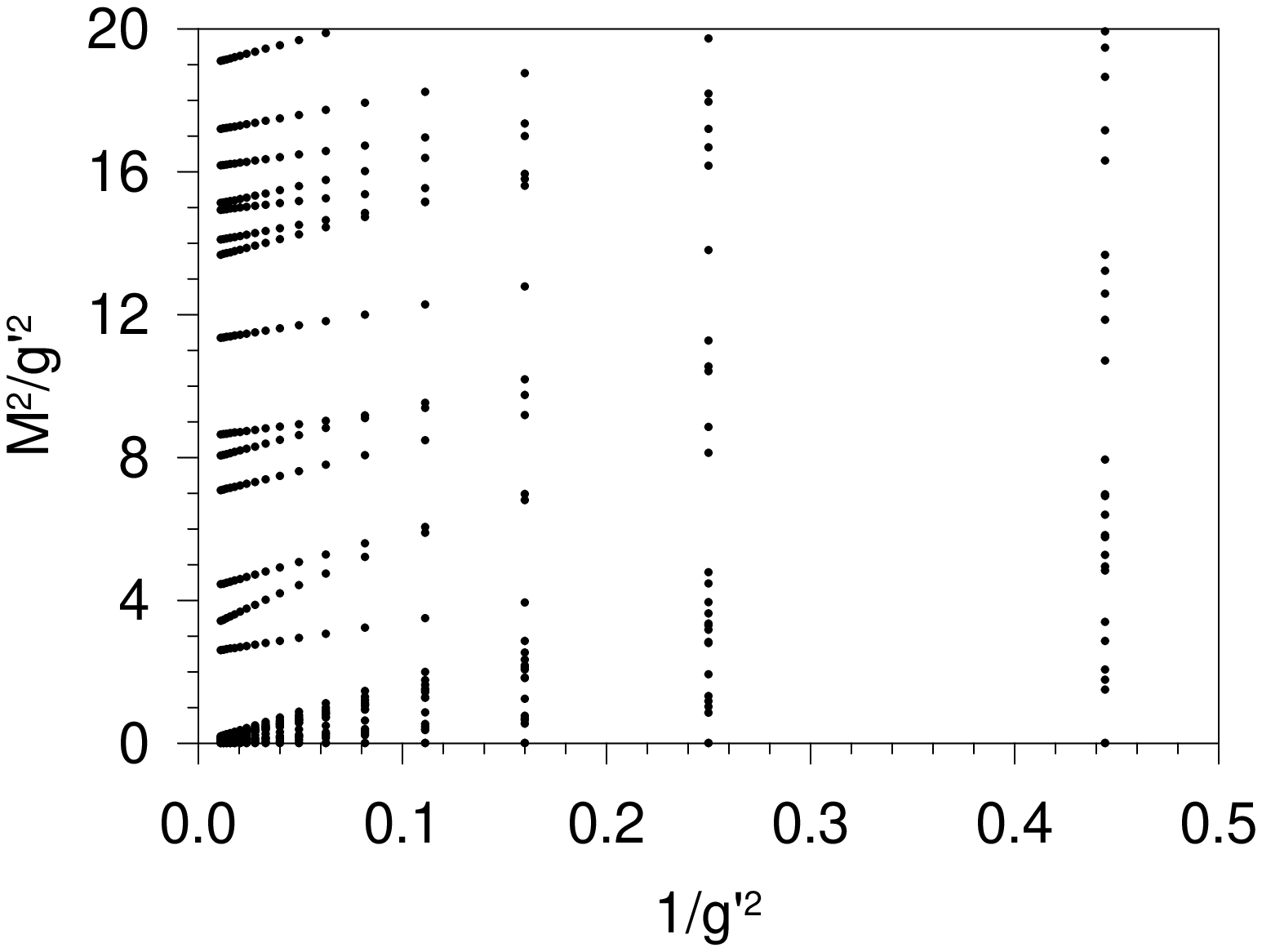,width=3.2in}  &
\psfig{file=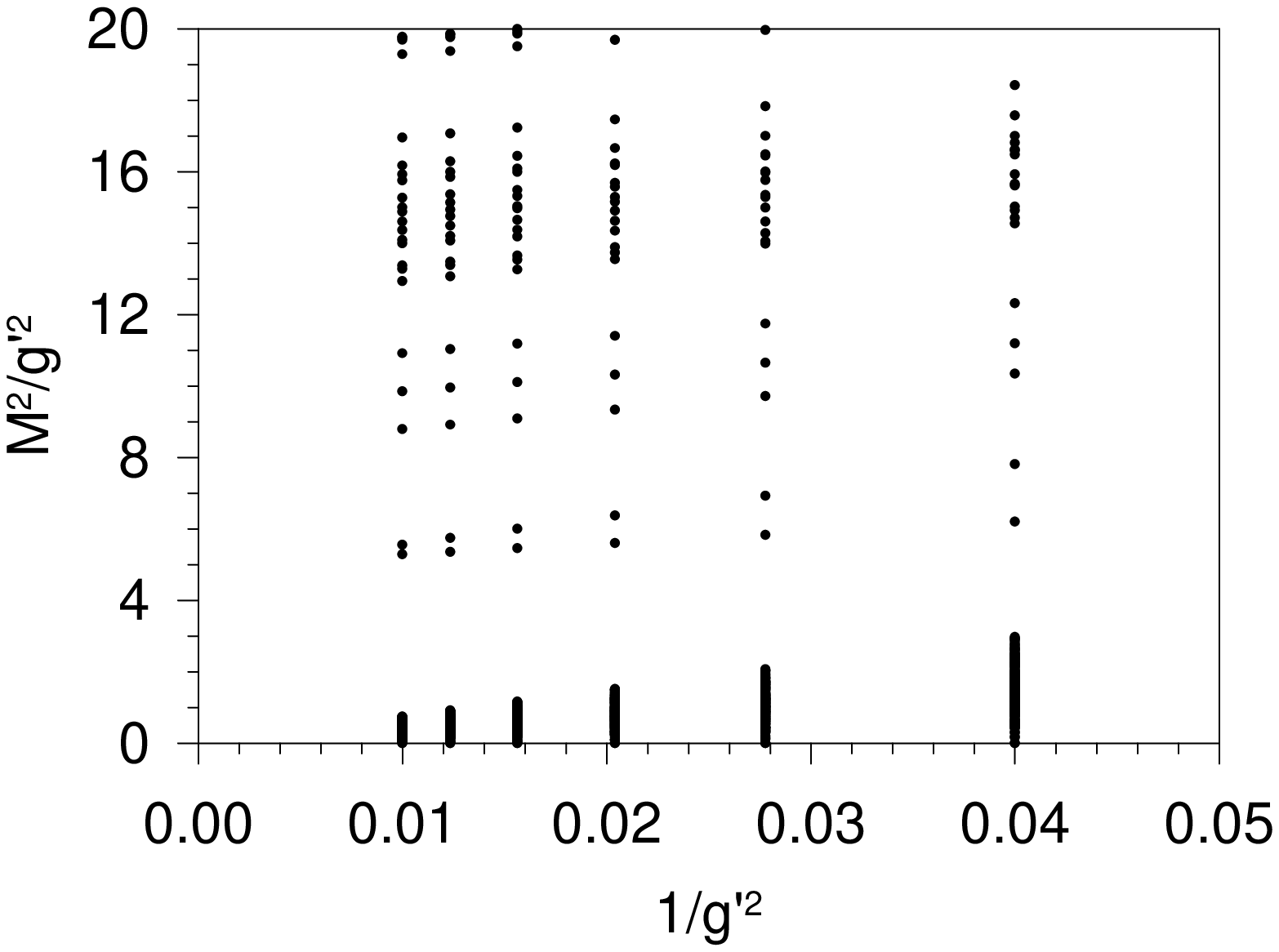,width=3.2in}  \\
(a) & (b) \\
\psfig{file=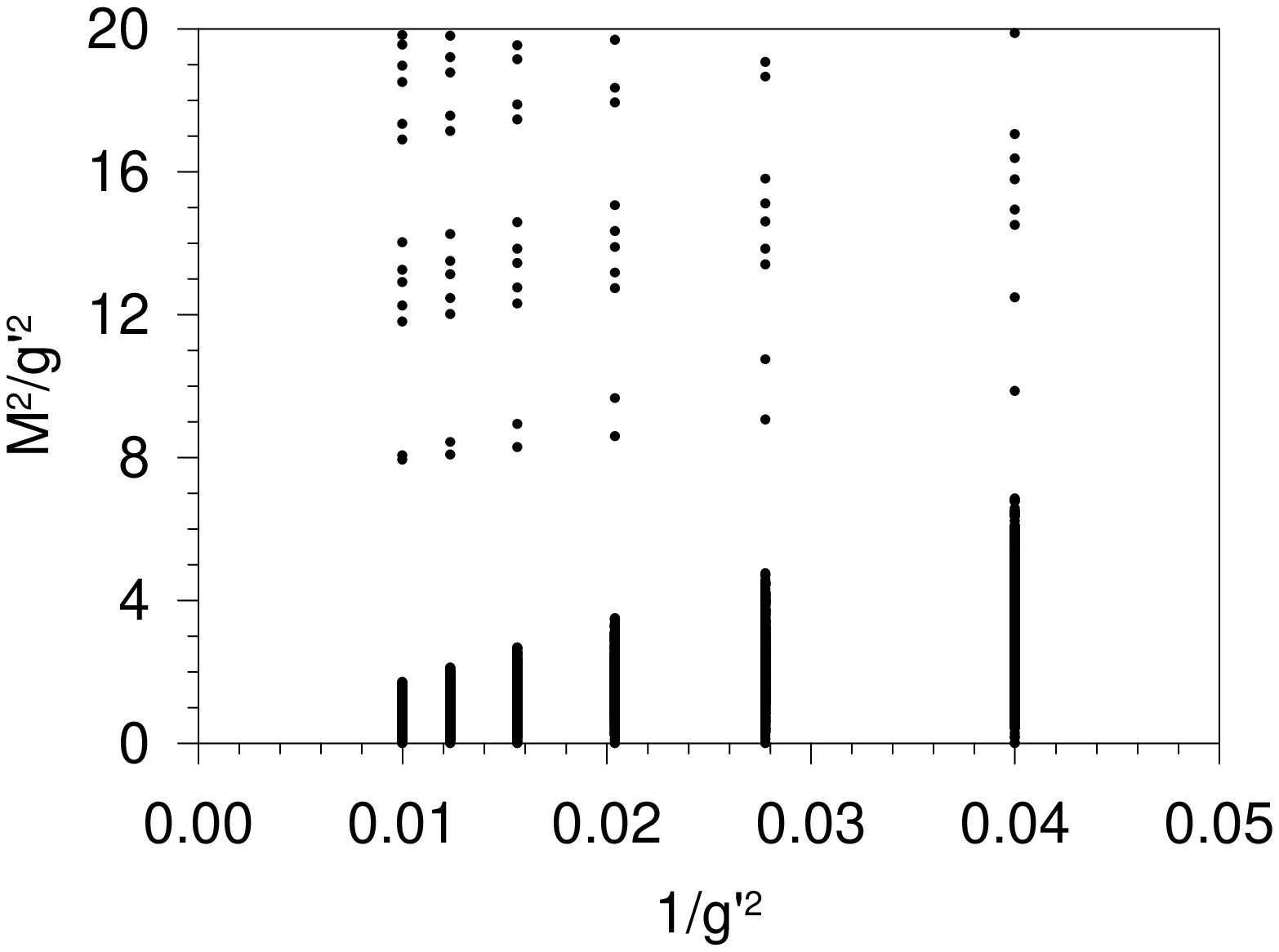,width=3.2in} &
\psfig{file=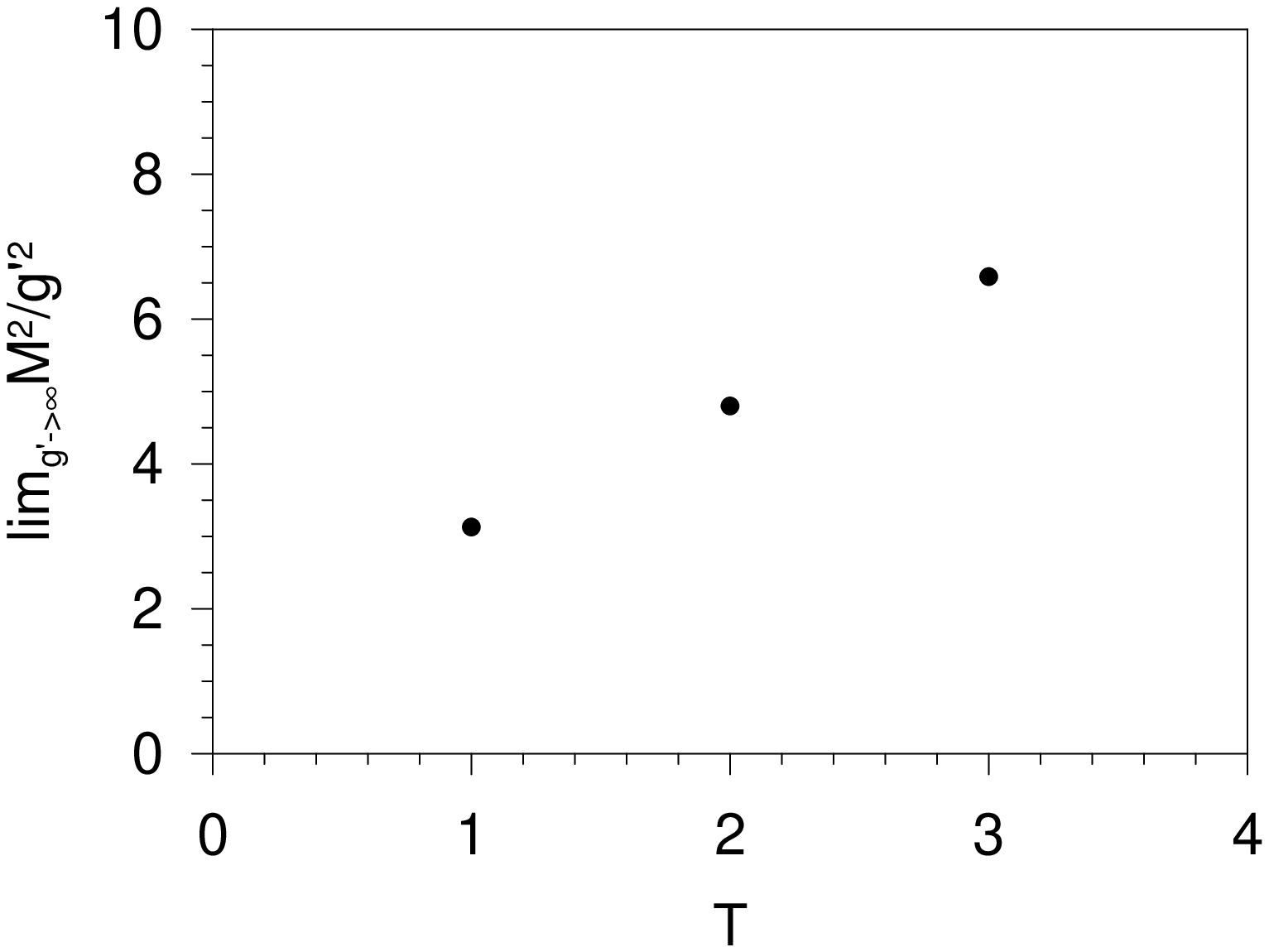,width=3.2in}  \\
(c) & (d) 
\end{tabular}
\caption{Bound-state masses squared $M^2$ in units of 
$4 \pi^2 / L^2$ divided by  $g'^2$ as functions of $1/g'^2$
in the symmetry sector $S=+1$, $P=+1$
for $K=5$ and (a) $T=1$, (b) $T=2$, and (c) $T=3$; and 
(d) the $T$ dependence of the intercepts of the
lowest-mass state in the strong-coupling sector of (a), (b), and (c).
\label{strong}}
\end{figure}

The Hamiltonian matrix is a quadratic matrix polynomial in $g'$,
and, therefore, it is natural to consider $M^2/g'^2$ as a function 
of $1/g'^2$. In Fig.~\ref{strong} we show plots for $K=5$ and $T=1$, 2 and 3. 
At the bottom of the strong-coupling band, several clear sets of separate 
masses appear.  They seem to move linearly as functions of $1/g'^2$, and 
it is natural to identify them as bound states.  From these three figures 
it is clear that these states have a strong $T$ dependence. 
In Fig.~\ref{strong}(d) we plot the intercept of a linear fit in 
$1/g'^2$ to the lowest state in the strong-coupling band. While we
only have three points, they clearly appear to have linear behavior in $T$. 

\begin{figure}[ht]
\centerline{\psfig{file=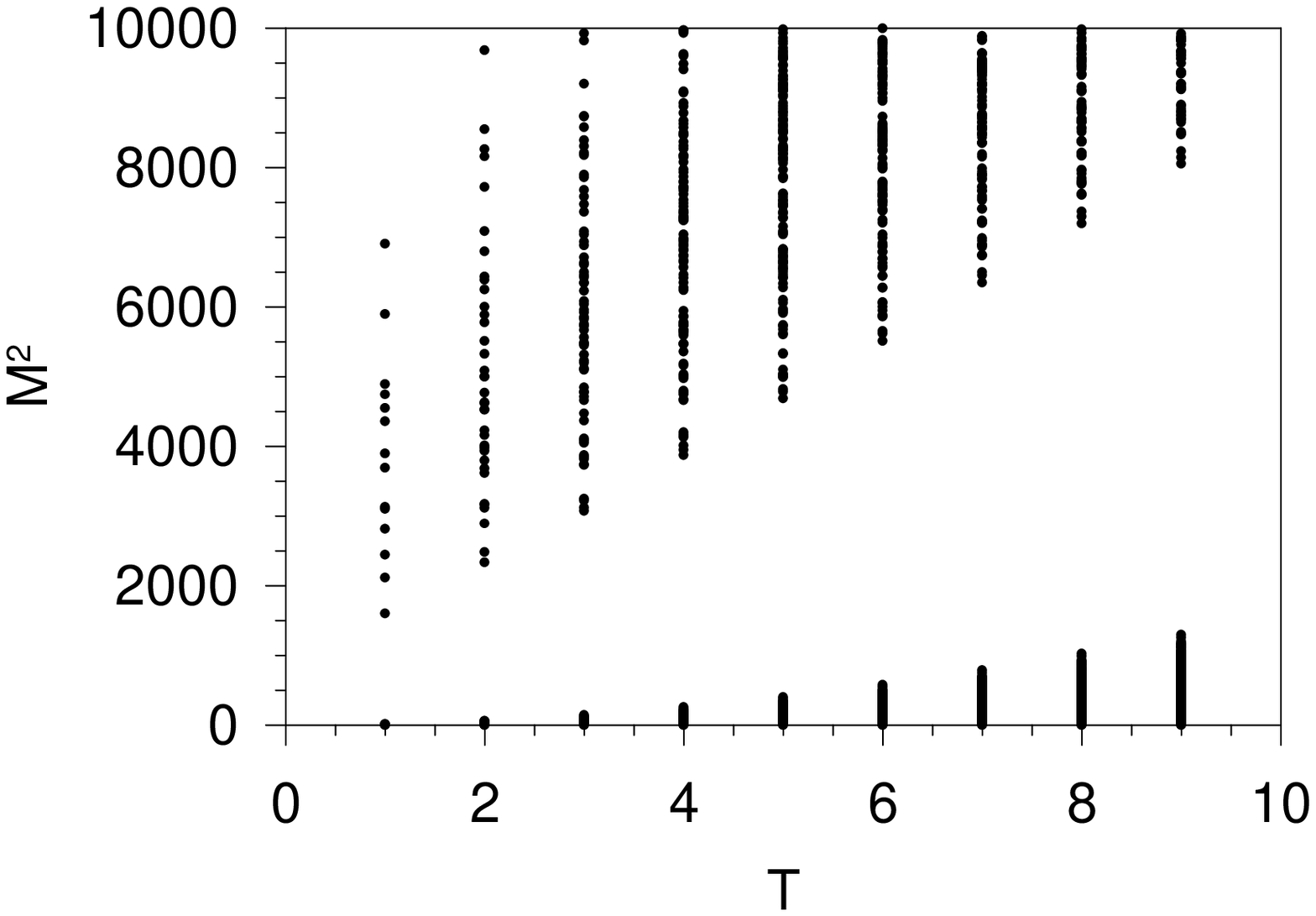,width=3.2in}} 
\caption{Bound-state masses squared
$M^2$ in units of $4 \pi^2 / L^2$ as functions of $T$, for
the symmetry sector $S=+1$, $P=+1$, longitudinal resolution $K=4$,
and coupling $g'=10$.
\label{fullT}}
\end{figure}

In fact, if we look at the full spectrum as a function $T$, as shown in
Fig.~\ref{fullT}, we see that the entire strong-coupling band appears to 
grow linearly with $T$, as the state we just discussed behaves. 
Given this behavior in $T$, it is not clear whether we should take these 
states seriously. If we view these states in terms of dimensionful 
quantities, the gauge coupling $g_{YM}$ and the transverse cutoff 
$\Lambda_\perp =2\pi T/L$, we see that the physical mass squared
${\cal M}^2$ is proportional to $g_{YM}^2 N_c \Lambda_\perp$. 
This is an expression 
for the bound-state mass that is independent of the length scale $L$; it is,
however, not what one expects for the continuum limit for the bound states 
of a finite theory with only one dimensionful parameter, $g_{YM}$. As we 
commented in~\cite{hhlpt99}, one would expect the continuum mass of such 
a theory to behave like $(g_{YM}^2 N)^2$.

\begin{figure}
\begin{tabular}{cc}
\psfig{file=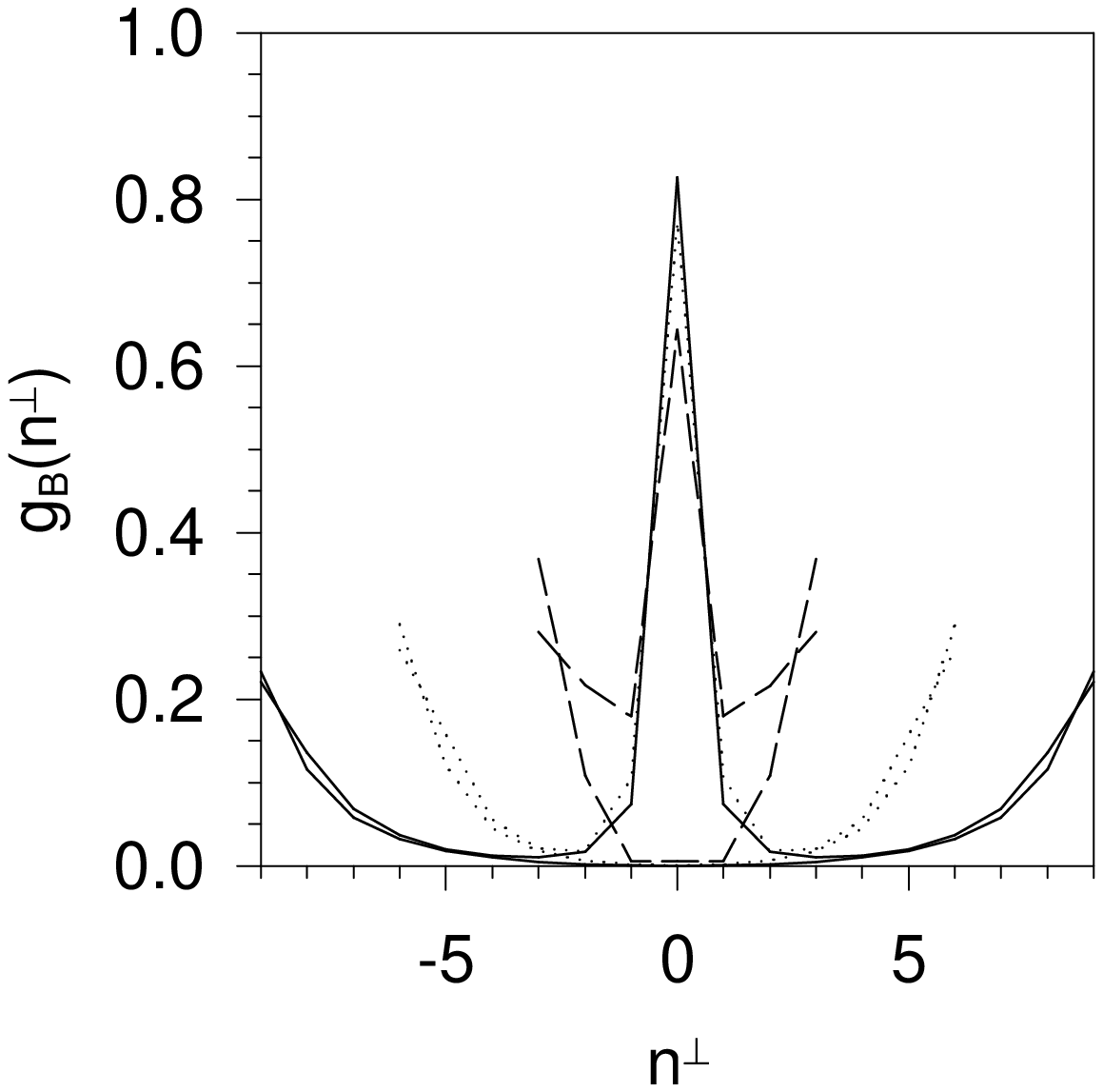,width=3.2in}  &
\psfig{file=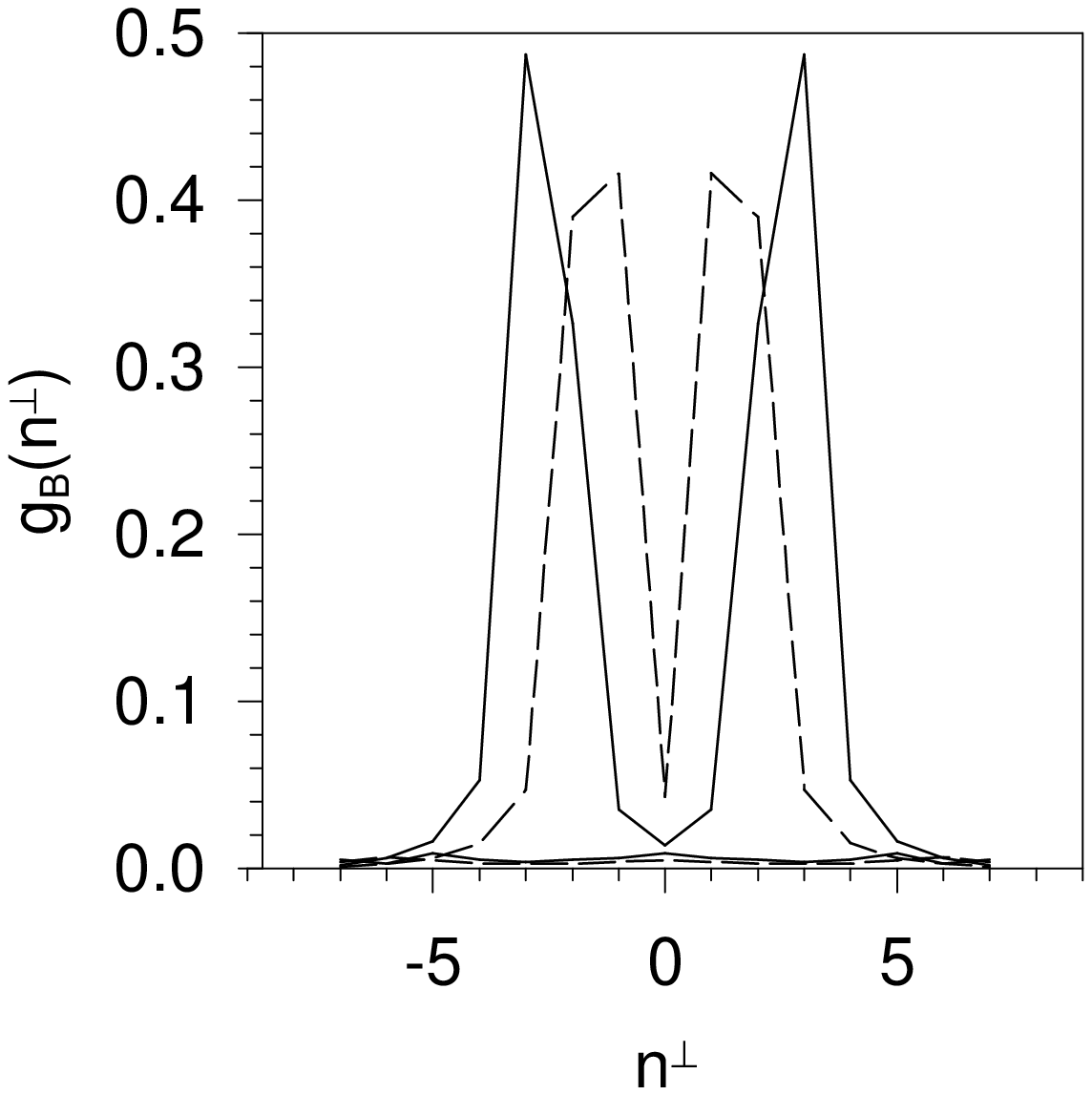,width=3.2in}  \\
(a) & (b)
\end{tabular}
\caption{Structure functions for some massive states
in the strong-coupling sector: (a) the lowest state in the sector, 
at $T=3,6,9$ 
(dashed, dotted, solid lines, respectively), and (b) two very massive states 
with $M^2=24,625$ (dashed line) and $M^2=22,728$ (solid line). 
For all cases, the symmetry sector is $S=+1$, $P=+1$, the longitudinal
resolution is $K=4$, and the coupling is $g'=10$.
\label{SFs4}}
\end{figure}

We know that this is a totally finite theory, so it would appear
that this behavior is very strange. It is interesting to
look at the structure functions for some of these states. In
Fig.~\ref{SFs4}(a) we show the structure function of the lowest
state in the strong-coupling band for various values of $T$.
Around zero transverse momentum we see in this figure a bound state similar to 
the bound states we saw in earlier structure functions. This
central peak converges quickly in $T$ and cannot be the main
reason for the $T$ dependence of this state. It rather comes from the
wings that appear at large transverse momentum.  As we increase $T$, we see a
larger and larger part of these bound states, and this gives rise
to the strong $T$ dependence of the total state. 

%
%

In Fig.~\ref{SFs4}(b) we show the structure functions of two other
states in the strong-coupling band. These states do not have the central peak
as the one in
Fig.~\ref{SFs4}(a), and the large wings that we saw before are replaced
by two bumps at non-zero momentum. It appears that the property that 
characterizes these states in the strong-coupling band is this multi-hump 
distribution function.

We now return to the unphysical states that we saw in the 
intermediate-coupling band. It is convenient to consider them at large 
coupling even though they are in the intermediate coupling band. The 
identifying property of these states is their low average parton number. 
When $K=4$ and $g'=10$, most states have $\langle n\rangle\simeq 4$, 
whereas these states have $\langle n\rangle\simeq 2$. 
The mass of the lowest of these states grows rapidly with 
$T$, similar to the states that we found in the strong-coupling band.

\begin{figure}
\centerline{\psfig{file=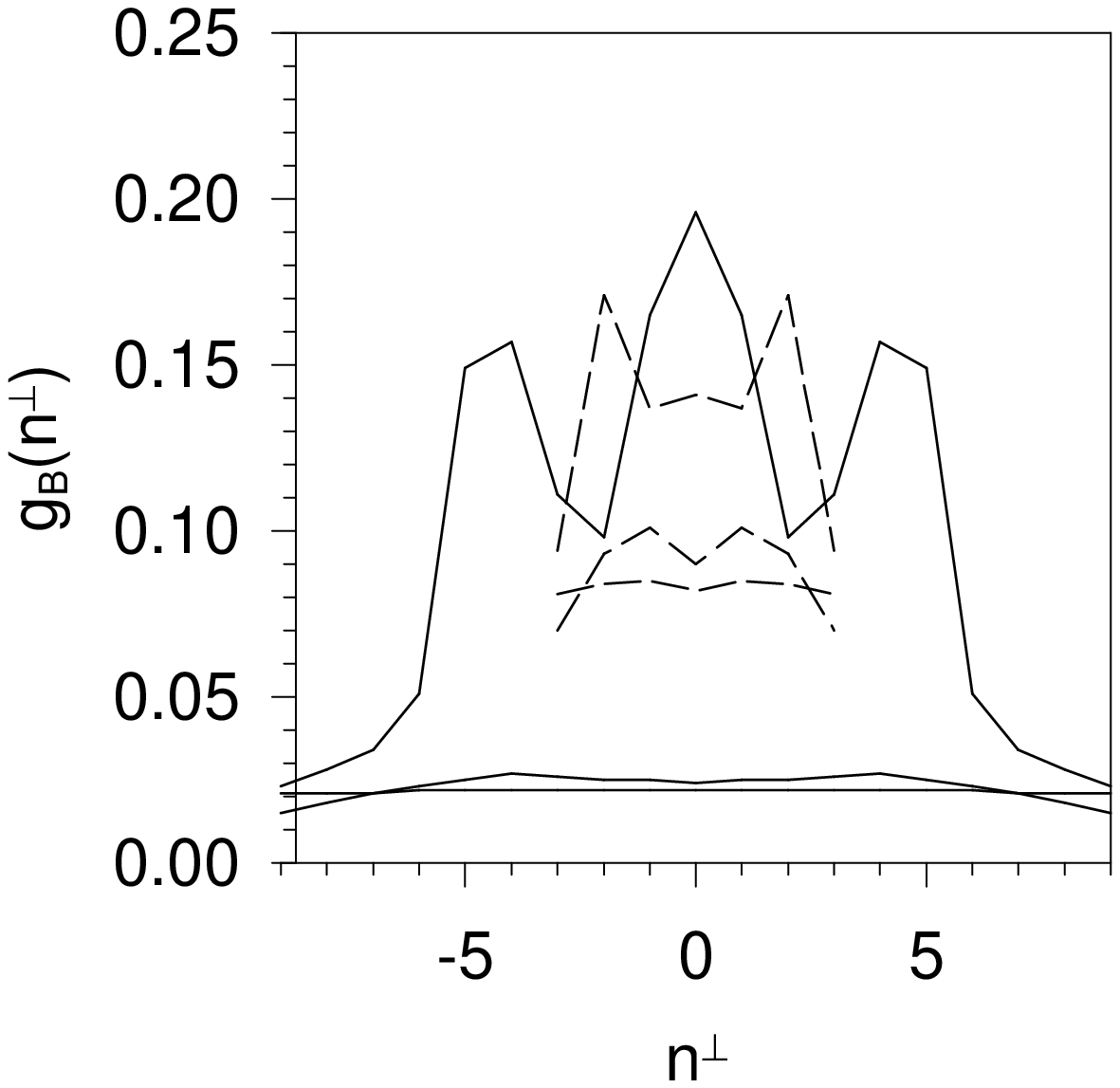,width=3.2in}}
\caption{Structure functions for an unphysical state.
Shown are the probabilities to find a boson 
with longitudinal momentum $n=1,2,3$ (each with decreasing
amplitude) for different cutoffs 
$T=3$ (dashed lines) and $T=9$ (solid lines).
The symmetry sector is $S=+1$, $P=+1$, the longitudinal resolution
is $K=4$, and the coupling is $g'=10$.
\label{SFs3}}
\end{figure}

The structure function for this state is shown in Fig.~\ref{SFs3}.  We see a 
shape that is somewhat similar to those for states in the strong-coupling band.
There now appear to be three bound states with all three peaks resolved 
but not cleanly separated. We see that, as the transverse resolution is 
increased from three to nine, the three peaks become much more distinct. 
There is no doubt that this 
change gives rise to the strong $T$ dependence. 
It is not clear at this time exactly why certain of these multi-hump 
distributions appear in the lower band and others appear in the upper band.

%
%

\begin{figure}[ht]
\begin{tabular}{cc}
\psfig{file=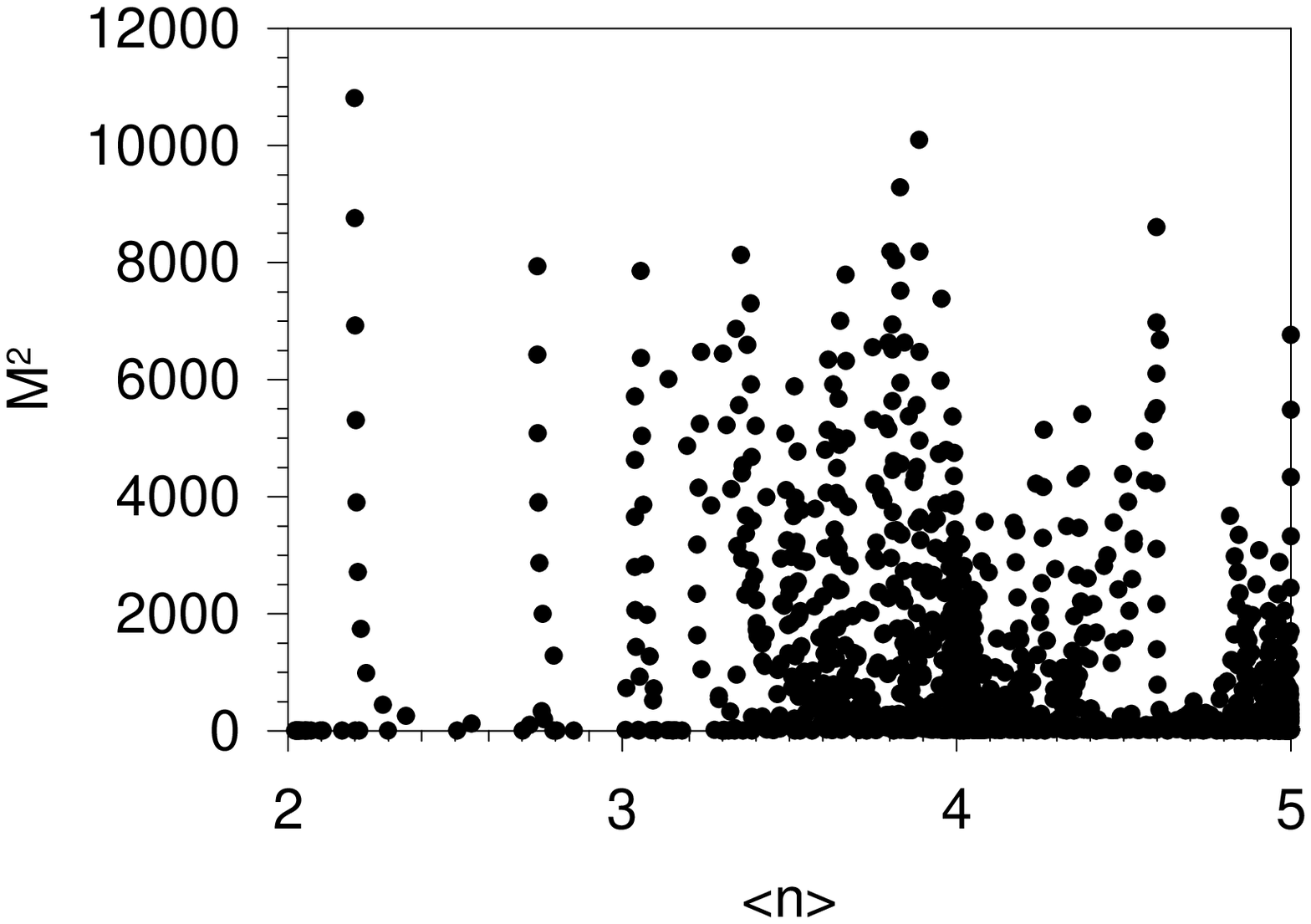,width=3.2in}  &
\psfig{file=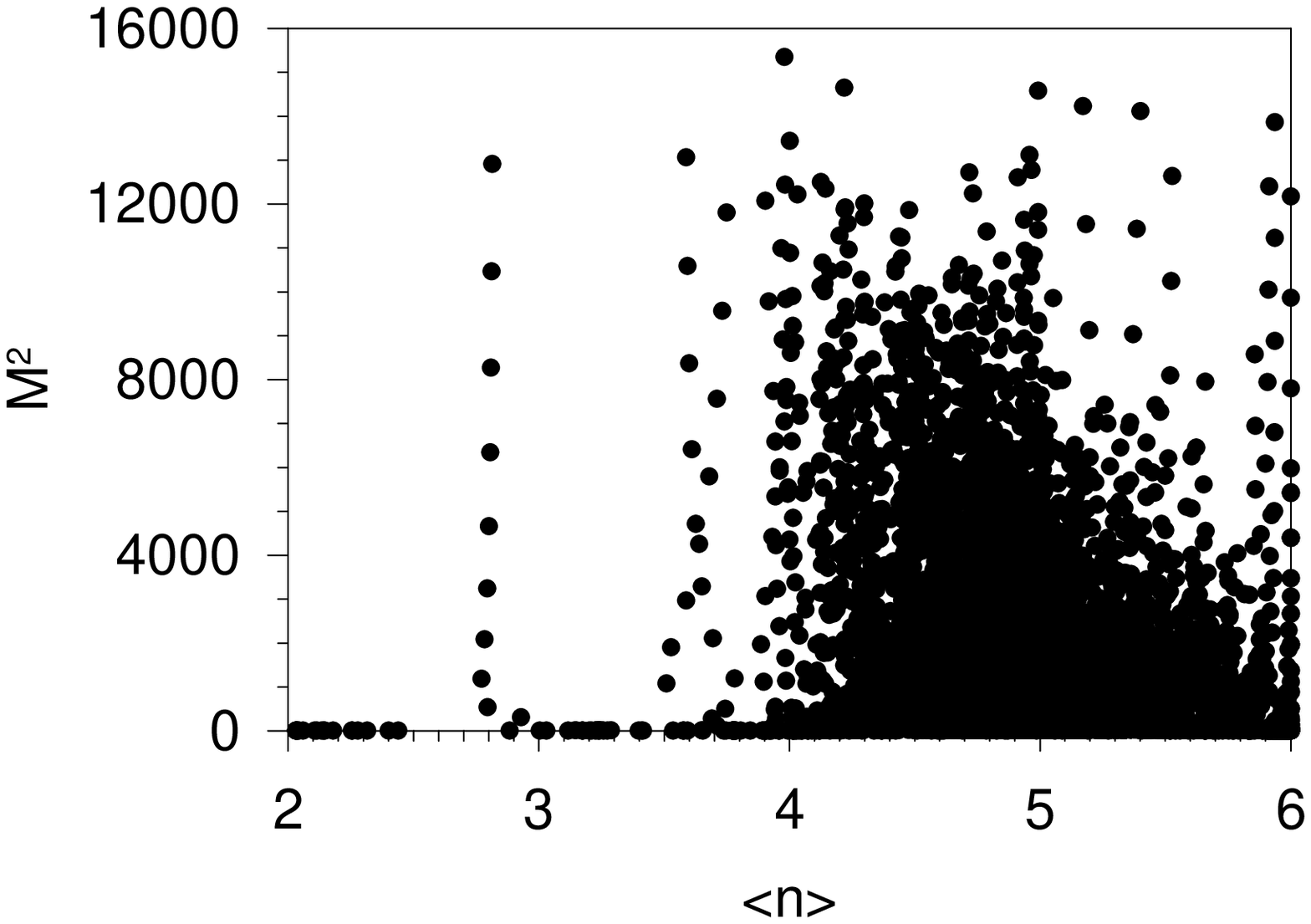,width=3.2in}  \\
(a) & (b)
\end{tabular}
\caption{Bound-state masses squared $M^2$ in units of $4 \pi^2 / L^2$ 
as functions of the average number of particles $\langle n\rangle$ for
(a) $K=5$, $T=1$ and (b) $K=6$, $T=1$.  Several different values of
the coupling are included; they are $g'=0.1$, 0.5, 1, 1.5, 2, 3, 4,\ldots,10.
The symmetry sector is $S=+1$, $P=+1$.
\label{averagen}}
\end{figure}
\begin{figure}[ht]
\begin{tabular}{cc}
\psfig{file=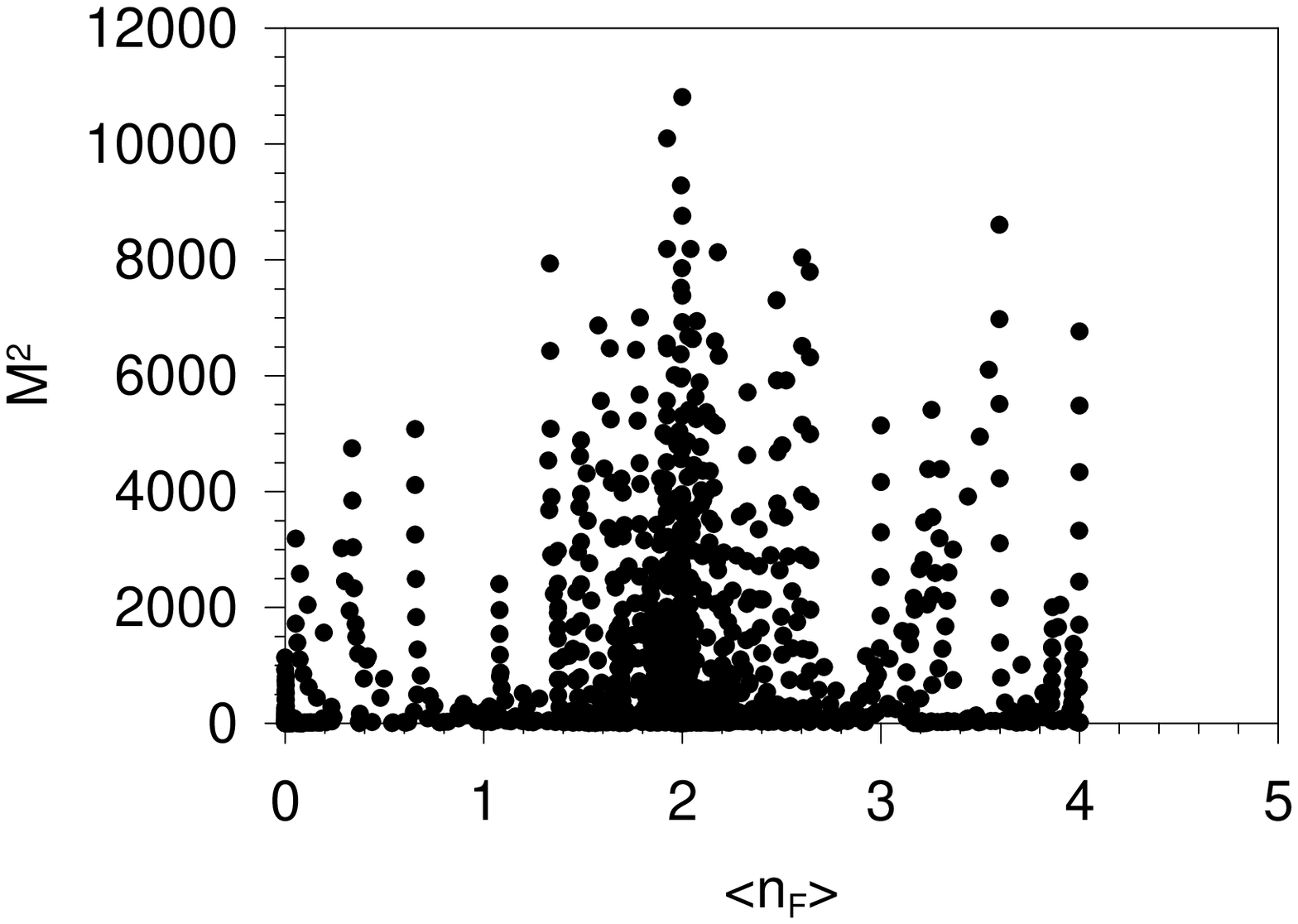,width=3.2in}  &
\psfig{file=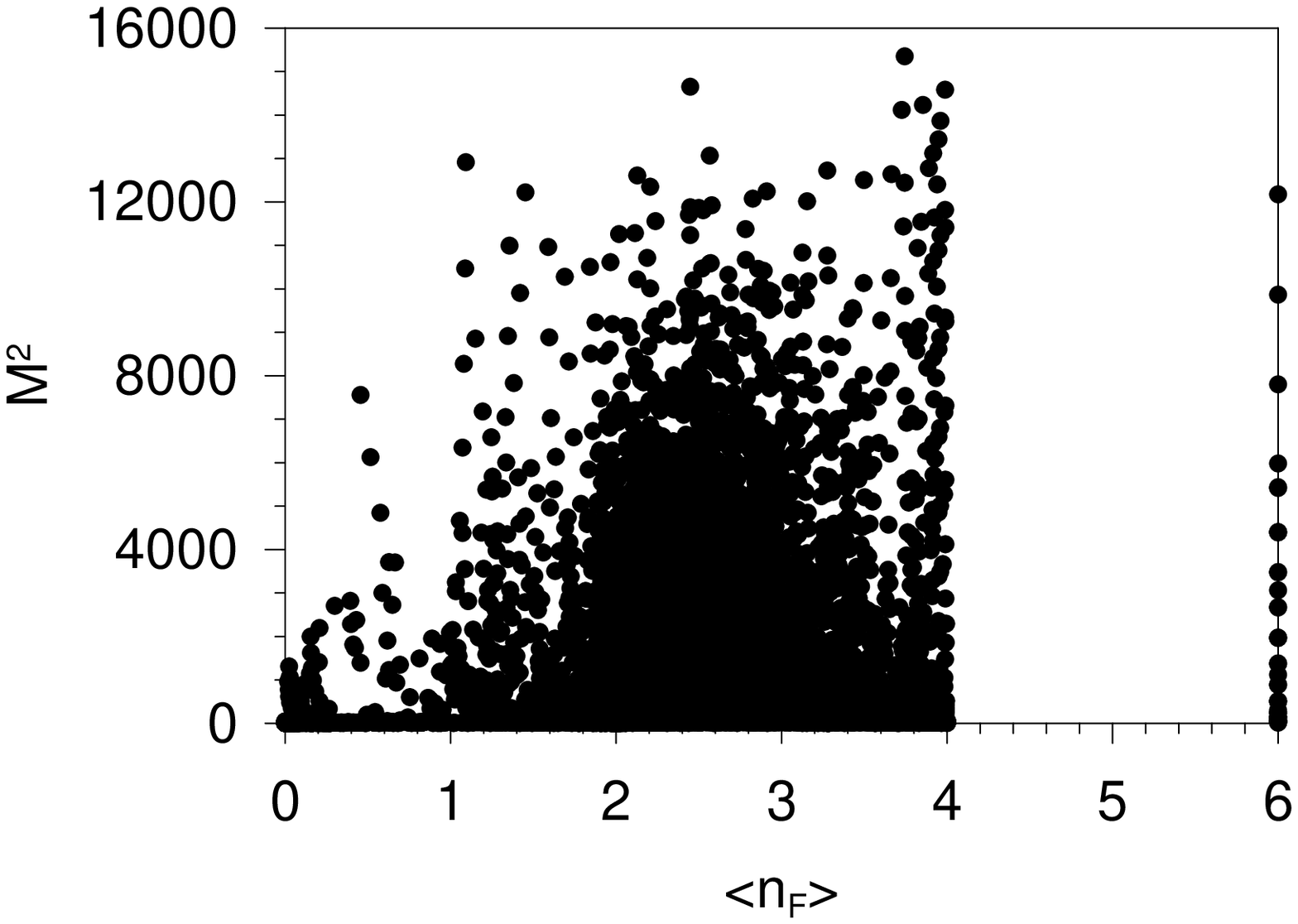,width=3.2in}  \\
(a) & (b)
\end{tabular}
\caption{Same as Fig.~\ref{averagen} but for masses as functions
of the average number of fermions $\langle n_F\rangle$.
\label{averagenf}}
\end{figure}

Let us now turn to some of the global properties of the full spectrum. 
In Fig.~\ref{averagen} we show the bound-state spectrum as a function 
of the average number of particles at $K=5$, $T=1$ and $K=6$, $T=1$. 
We superimpose the spectra for $g'=0.1$, 0.5, 1, 1.5, 2, 3, 4,\ldots,10. 
As we noted previously, the density clusters near the maximum value
allowed for $\langle n\rangle$, which is $K$. In addition we see sets of 
trajectories with increasing masses at small $\langle n \rangle$. In fact,
these trajectories with the low $\langle n \rangle$ correspond to the 
most massive states in the entire spectrum. 

In Fig.~\ref{averagenf} 
we show the bound-state spectrum as a function of the average
number of fermions. In spite of the fact that we are dealing with a 
supersymmetric theory, the average number of fermions behaves quite 
differently than that of the bosons. It does not appear to grow nearly as fast. 
In the intermediate-coupling region we saw that the number of 
fermions was nearly a sharp quantum number in the bound state,
that is $\langle n_F^2\rangle =\langle n_F\rangle^2$. In comparing 
the figures for $K=5$ and $K=6$, in Fig.~\ref{averagenf}, we see 
a particularly striking feature: there are no states in the range from
4 to 6 for $K=6$. That is, all the bound states with six fermions have
exactly six fermions. The interpretation of this fact is not yet clear.

\section{Discussion}
\label{summary}

In this paper we present a first-principles calculation of the massive 
spectrum and the wave functions of ${\cal N}=1$ SYM theory in 2+1 dimensions. 
This is the first such calculation and will provide a benchmark for 
future calculations using lattice and other methods. The calculation 
was performed using SDLCQ, which has been successfully applied to a number of 
(1+1)-dimensional theories. The success of this method stems from
its retention of supersymmetry at every step of the numerical approximation. 

In our formulation this theory has three exact symmetries: supersymmetry, 
parity $P$, and an orientation symmetry we call $S$. These are all $Z_2$ 
symmetries, and they allow us to reduce the size of the matrices in the 
numerical approximation by a factor of 8. Supersymmetry and 
parity give rise to a four-fold degeneracy which is now split between
different symmetry sectors.  The elimination of this four-fold
degeneracy simplifies the application of the Lanczos diagonalization
algorithm.
 
The theory has three dimensionful constants with the dimension 
of (length)$^{-1}$: $g_{YM}^2N_c$; $\Lambda_\perp$, the transverse momentum 
cutoff; and $1/L$, the reciprocal of the length of the compact dimension. 
We find spectra that behave as $1/L^2$, $g^2_{YM} N_c \Lambda_\perp$, and 
$\Lambda^2_\perp$. In Fig.~\ref{full} we see that the spectrum divides 
into three regions, a weak-coupling region and two bands at higher coupling. 
The weak-coupling region is discussed in detail in 
Ref.~\cite{hhlpt99}. There we found massless states and very light states 
that are totally determined by the (1+1)-dimensional theory, as well as 
heavier states whose masses come primarily from the transverse 
momentum of the constituents. These states all scale like $1/L^2$. 

Of the two bands at higher coupling, we call the lower one the
intermediate-coupling band. Most of the 
bound states in this band are dominated by a very large number of constituents 
as the coupling gets strong. This fact limits the region in coupling space 
that we can explore using SDLCQ. We have explored this region 
up to an intermediate 
coupling, and we find a well defined spectrum which converges very rapidly 
as we increase the transverse momentum cutoff. We also found that the 
transverse momentum distributions converge exceptionally well with
transverse momentum cutoff. Convergence in the longitudinal resolution 
is not as rapid as the transverse convergence, but it still appears to 
be well behaved.  We extrapolate to infinite resolution in both the 
longitudinal and transverse directions and present a list of some 
of these states and their properties. 
The masses of all of these states scale like $1/L^2$. It is interesting to 
note that, while the average number of particles grows rapidly with the 
coupling, this is not true for the average number of fermions. Typically 
the states we studied in this band have precisely either zero fermions or 
two fermions. In fact, this is a property that persists throughout the 
full spectrum. It is as if each bound state has a valence number 
of adjoint fermions that characterizes the state.

In addition to these states, there are other states in the intermediate-coupling
band that we discussed previously~\cite{alp99b,hhlpt99} and have called 
unphysical states. The distinctive properties of these states are that the 
average number of particles is small and does not appear to saturate the 
resolution. The transverse momentum distributions are multi-humped, and the 
masses grow like $\Lambda^2_\perp$ and fall with increasing coupling. 
Therefore, as we take the transverse cutoff to infinity, these states 
appear to decouple. It could be of some importance to find a physical 
meaning for these states, because they have a small average
number of particles, as one might expect in a QCD-like theory, but even if 
their masses were to stabilize at some large transverse cutoff beyond what we 
currently can reach, they would be so massive that it is hard to see how they 
could be relevant. The only hope might be that at some very large coupling 
they fall back into an interesting region.

In the strong-coupling band we found states that scaled like $g^2_{YM} N_c$ at
large coupling. We were motivated to look for this behavior because the 
supercharge is linear in the coupling. These states  
have multi-hump transverse momentum distributions, and their  masses 
go like $g^2_{YM} N_c \Lambda_\perp$. We saw that some of the transverse 
momentum distributions have two clearly separated and symmetric humps. 
It is as if half of the constituents in these bound states are going one 
way while the other half are going the other way, leaving no constituents
that are at rest relative to the center of mass of the bound state. 
It almost appears as though we are looking at two bound 
states~\cite{anp98,ghk97}. If this were the case we would expect to see 
the same bound states elsewhere in the spectrum with a different relative 
transverse momentum and a different total energy.  We have looked for 
these states but could not find them. Also, this picture does not naturally 
seem to lead to an explanation of the linear growth of the mass with the 
transverse momentum cutoff.

We have looked at the global properties of this strongly coupled band and found
unusual behavior as a function of the average number of fermions and of the 
average number of particles. Numerical limitations unfortunately make this 
band more difficult to investigate, and, while it remains possible that 
interesting physics might emerge from these states, the most likely result is 
that they decouple from the physical spectrum. We must note that we do not find 
states that scale as $g^4_{YM} N^2_c$, which is the dependence one would 
expect in $R^3$, where the coupling is the only dimensionful parameter 
in this finite theory.

From what we have learned about the spectrum of this theory, there appear to be
three natural directions for future investigations. One is to explore the fact 
that all these states seem to prefer a small valence number of 
adjoint fermions. We might consider a supersymmetric-like theory which 
only has adjoint fermions. Such a theory is obtained in 1+1 dimensions by 
simply dropping the boson terms in the supercharge~\cite{anp98,kutasov93}. 
One could try the same thing here.  Alternatively, one can decouple the 
bosons by adding a mass term for them.  These are two simple extensions
of the present work that might avoid the large number of adjoint bosons 
in the bound states. The third alternative is to consider a Chern--Simons 
extension of this theory. This has the advantage of maintaining exact 
supersymmetry and giving the constituents a mass. In addition to these
three directions for the ${\cal N}=1$ theory, we can consider the next class of 
interesting models by adding more supersymmetry. The ${\cal N}=2$ theory
would be particularly interesting because it is the dimensional reduction of
the ${\cal N}=1$ theory in 3+1 dimensions. 

\acknowledgments
This work is supported in part by the US Department of Energy.
J.R.H. thanks the Department of Physics of the Ohio State 
University for its hospitality during a visit there while this work
was being completed.


\end{document}